\documentclass[%
 reprint,
 amsmath,amssymb,
 aps,
]{revtex4-1}

\usepackage{graphicx}
\usepackage{dcolumn}
\usepackage{bm}
\usepackage{xcolor}

\begin{document}

\preprint{AIP/123-QED}

\title{Understanding mechanisms of thermal expansion in PbTiO$_3$ thin-films from first principles: role of high-order phonon-strain anharmonicity}

\author{Ethan T. Ritz}
\affiliation{ 
Sibley School of Mechanical and Aerospace Engineering, Cornell University, Ithaca, New York 14853, USA}%
\author{Nicole A. Benedek}
\email{nbenedek@cornell.edu}
\affiliation{%
Department of Materials Science and Engineering, Cornell University, Ithaca, New York 14853, USA
}%

\date{\today}

\begin{abstract}
The thermal properties of materials are critically important to various technologies and are increasingly the target of materials design efforts. However, it is only relatively recent advances in first-principles computational techniques that have enabled researchers to explore the microscopic mechanisms of thermal properties, such as thermal expansion. We use the Gr\"{u}neisen theory of thermal expansion in combination with density functional calculations and the quasiharmonic approximation to uncover mechanisms of thermal expansion in PbTiO$_3$ thin-films in terms of elastic and vibrational contributions to the free energy. Surprisingly, we find that although the structural parameters of PbTiO$_3$ thin-films evolve with temperature as if they are dominated by linear elasticity, PbTiO$_3$ thin-films are strongly anharmonic, with large changes in the elastic constants and Gr\"{u}neisen parameters with both misfit strain and temperature. We show that a fortuitous near-cancellation between different types of anharmonicity gives rise to this behavior. Our results illustrate the importance of high-order phonon-strain anharmonicity in determining the temperature-dependent structural parameters of PbTiO$_3$ thin-films, and highlight the complex manner in which thermal expansion, misfit strain and elastic and vibrational properties are intertwined.

\end{abstract}

\pacs{Valid PACS appear here}
\keywords{Suggested keywords}
\maketitle
\section{Introduction}
Recent advances in theoretical and experimental techniques have promoted increasing interest in the phonon and thermal properties of materials by making it possible to more easily access the information required to elucidate atomic-scale mechanisms.\cite{tucker2005negative, mounet05, kim2018nuclear, ritz2018interplay} Particularly exciting progress has been made in the family of metal-organic frameworks, where it now not only appears possible to design materials with unconventional thermal responses (such as negative thermal expansion), but also to tune those responses in a highly controllable way \cite{goodwin08,goodwin08b,goodwin09}. In contrast, negative thermal expansion is both rarer and apparently more difficult to tune in inorganic framework materials, such as complex oxides \cite{lind2012two,miller2009negative, barrera2005negative}. The most common approach is chemical substitution, although it is generally challenging to change either the sign or the magnitude of the thermal expansion significantly in this way (recent work provides an impressive exception \cite{yanase2011effects, ablitt2019tolerance}). A significant barrier to designing effective strategies for highly tunable thermal expansion is that, with few exceptions, the microscopic factors controlling the temperature evolution of structural parameters are poorly understood.

The ferroelectric $P4mm$ tetragonal phase of PbTiO$_3$ exhibits negative volumetric thermal expansion from approximately room temperature up to 760 K, where it undergoes a structural phase transition to a cubic $Pm\bar{3}m$ phase \cite{chen2005thermal,shirane1951phase}. Thermal expansion in the ferroelectric phase is characterized by a shrinking $c$-axis and an expanding $a$-axis -- the $c$-axis contracts faster than the $a$-axis expands, so the net effect is a decrease in volume with increasing temperature. Interestingly, recent theoretical work has shown that negative thermal expansion in PbTiO$_3$ is actually \emph{driven} by the expanding $a$-axis \cite{ritz2018interplay}. However, strong elastic (Poisson-like) coupling between the $a$- and $c$-axes pulls the $c$-axis down as the $a$-axis expands.

It is possible to grow high-quality thin-films of PbTiO$_3$ on various substrates, and a wealth of temperature-dependent structural and ferroelectric polarization data are available in the literature. Thin-film systems are excellent platforms for probing mechanisms of thermal expansion because different substrates will impart different misfit strains (which will change with temperature, depending on the thermal expansion coefficient of the substrate\cite{janolin2007temperature,janolin09}) and hence different elastic boundary conditions to the growing film. Hence, it becomes possible to understand how changes in the unit cell dimensions affect thermal expansion behavior, all while keeping the chemical composition constant.

In previous work, we used first-principles density functional theory (DFT) and the quasiharmonic approximation (QHA) to investigate the thermal expansion of PbTiO$_3$ thin-films on SrTiO$_3$, DyScO$_3$ and (La$_{0.29(5)}$Sr$_{0.71(5)}$)$_{\mathrm{A-site}}$(Al$_{0.65(1)}$Ta$_{0.35(1)}$)$_{\mathrm{B-site}}$O$_3$ (LSAT) substrates. Our key findings were that: 1) the misfit strain does not just change the properties of PbTiO$_3$ thin-films, it changes how they change with temperature, and 2) the mismatch in thermal expansion coefficients between PbTiO$_3$ and the substrates had a striking effect on thin-film properties. For a given misfit strain, we found qualitative changes in thermal expansion behavior and ferroelectric transition temperatures depending on the thermal expansion coefficient of the substrate. 

In this work, we use the Gr\"{u}neisen theory of thermal expansion in combination with DFT calculations and the quasiharmonic approximation to elucidate the microscopic mechanisms underlying the structural properties of PbTiO$_3$ thin-films in terms of elastic and vibrational contributions to the lattice parameters at finite temperatures. Surprisingly, we show that the $c$-axis lattice parameter (perpendicular to the substrate) evolves with temperature as if it is almost entirely dominated by linear elastic effects, despite the fact that PbTiO$_3$ thin-films exhibit the hallmarks of strong anharmonicity. We find that this behavior is due to a fortuitous near-cancellation between the contributions to thermal expansion from anharmonic elasticity and from vibrational thermal stress, the details of which depend both on the mismatch between the lattice parameters of the film and substrate, as well as the mismatch between their rates of thermal expansion. We use our findings to explain the temperature evolution of the structural, elastic, vibrational and thermal expansion properties of PbTiO$_3$ thin-films on experimentally available substrates.

\section{\label{sec:Methods}First-Principles Calculations}
All calculations were performed using density functional theory, as implemented in Quantum Espresso 6.5.0. \cite{giannozzi2009quantum} We used the Wu-Cohen exchange-correlation functional \cite{WC2006} with Garrity-Bennett-Rabe-Vanderbilt ultrasoft pseudopotentials.\cite{garrity2014pseudopotentials} The following states were included in the valence for each element: 5d$^{10}$6s$^2$6p$^2$ for Pb, 3s$^2$3p$^6$4s$^2$3d$^1$ for Ti, and 2s$^2$2p$^4$ for O. Zero temperature unit cell lattice parameters and atomic positions of $P4mm$ PbTiO$_3$ were converged with respect to the plane wave cutoff energy and $\mathbf{k}$-point mesh density to within 0.001 \AA. Unless otherwise mentioned, structural parameters were found to be converged at a force cutoff threshold of $3.0 \times 10^{-5}$ Ry/bohr using a 6 $\times$ 6 $\times$ 6 Monkhorst-Pack (MP) mesh and a 100 Ry plane wave cutoff energy, compared with MP meshes up to 12$\times$12$\times$12 and plane wave cutoffs up to 120 Ry. Phonon dispersion calculations were performed using density functional perturbation theory on an 8 $\times$ 8 $\times$ 8 $\mathbf{q}$-point grid.

Finite temperature structural parameters and elastic constants were predicted using a quasiharmonic approximation (QHA) to the Helmholtz free energy -- in this study, we closely follow the framework outlined in Refs. \citenum{ritz2019thermal} and \onlinecite{ritz2020strain} regarding the application of the QHA. Our grid of strained systems consisted of a 6 $\times$ 12 grid spanning -1.5 to +2.5\% strain along the $a$-axis of the $P4mm$ tetragonal phase, and -4.0\% to +5.1\% strain along $c$, augmented with 20 additional points with strains spanning -1.1\% to -0.3\% strain in $a$ and -4.0\% to +5.1\% strain in $c$. These points were chosen to ensure good convergence of finite-temperature elastic constants and to make sure that for bulk PbTiO$_3$ and all strained films studied, the lattice parameters as a function of temperature would be bounded by these values \cite{instab}. Some structures at very large values of strain exhibited phonons with imaginary frequencies; these points were discarded from the data set. Note that the strain values above are defined with respect to the 0 K lattice parameters of the $P4mm$ phase, including zero-point energy corrections from vibrational degrees of freedom (we find $a$ = 3.891~\AA~and $c$ = 4.174 \AA). 

We use the same definitions of misfit strain and thermal expansion mismatch as in Ref. \onlinecite{ritz2020strain}; we reproduce the definitions here for convenience. For the calculations of epitaxially strained PbTiO$_3$, we clamp the in-plane ($a$ and $b$) lattice parameters of PbTiO$_3$ to the temperature-dependent lattice parameters of the substrate. That is, our misfit strain is defined as,
\begin{equation}
    \varepsilon_a(T) = \frac{a_{\mathrm{substrate}}(T) - a_{\mathrm{PTO}}(T)}{a_{\mathrm{PTO}}(T)},
    \label{eq:misfit}
\end{equation}
where $a_{\mathrm{substrate}}(T)$ is the lattice parameter of the substrate at some temperature $T$, and $a_{\mathrm{PTO}}(T)$ is the in-plane lattice parameter of bulk PbTiO$_3$ at the same temperature from our QHA calculations. We can re-write this (as in Ref. \citenum{ritz2020strain}) as

\begin{equation}
    \varepsilon_a(T) = \varepsilon^{300}_a+\varepsilon^{thermal}_a(T),
    \label{eq:misfit_2}
\end{equation}

where $\varepsilon^{300}_a$ is the strain at a reference temperature of 300 K (the misfit strain is usually defined at room temperature in the thin-film literature by convention), and the additional strain that arises from changes in temperature $\varepsilon_a^{\mathrm{thermal}}$ is defined as
\begin{equation}
\begin{split}
\label{eq:thermal}
 \varepsilon_a^{\mathrm{thermal}}(T)=\int_{\tau=300}^T  \alpha_a(\tau)-\alpha_a^{\mathrm{bulk}}(\tau) \mathrm{d}\tau \\ = \int_{\tau=300}^T  \Delta \alpha_a(\tau) \mathrm{d}\tau,
\end{split}
\end{equation}

The misfit strain rate is defined as $\Delta \alpha_a(T)= \alpha_a(T)-\alpha_a^{\mathrm{bulk}}(T)$, where $\alpha_a(T)$ is the thermal expansion coefficient of the substrate, and $\alpha^{\mathrm{bulk}}_a(T)$ is the rate of thermal expansion along $a$ for unconstrained bulk PbTiO$_3$ (since we are considering PbTiO$_3$ films coherently clamped to the substrate, the thermal expansion coefficient of PbTiO$_3$ films along the $a$-axis is just that of the substrate).

\section{\label{analysis} Gr\"{u}neisen Theory of Thermal Expansion}
\begin{figure}
    \centering
    \includegraphics[width=9cm]{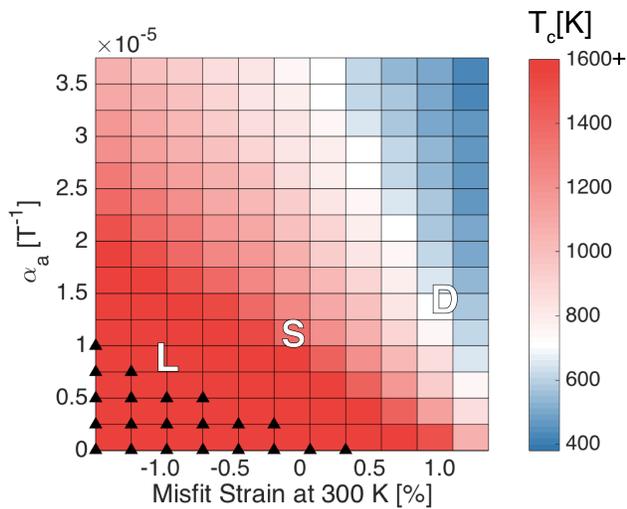}
    \caption{Variation in ferroelectric transition temperature and $c$-axis behavior of ferroelectric PbTiO$_3$ thin-films as a function of misfit strain and substrate coefficient of thermal expansion ($\alpha_a$) from our first-principles QHA calculations reported in Ref. \onlinecite{ritz2020strain}. The color chart for $T_c$ has white set to the bulk PbTiO$_3$ transition temperature of 760 K. Blue (red) squares indicate a strain and $\alpha_a$ combination that produces films with a lower (higher) $T_c$ than bulk. Triangles indicate combinations of strain and $\alpha_a$ that produce films in which the $c$-axis continues to grow with temperature and $T_c$ is suppressed. All other combinations of strain and $\alpha_a$ produce films in which the $c$-axis shrinks with temperature and $T_c$ is finite. Note that the films with the highest transition temperatures may exceed 1600 K or even be completely suppressed. The letters `L', `S' and `D' denote the strain and $\alpha_a$ conditions corresponding to growth on LSAT, SrTiO$_3$ and DyScO$_3$ substrates. See Ref. \onlinecite{ritz2020strain} for further details.}
    \label{tc_plot}
\end{figure}

In our previous work\cite{ritz2020strain} on thermal expansion in ferroelectric PbTiO$_3$ thin-films, we showed that the temperature of the transition ($T_c$) to the cubic $Pm\bar{3}m$ phase  depends \emph{qualitatively} not only on the misfit strain but also on the thermal expansion coefficient of the substrate ($\alpha_a$). For example, Figure \ref{tc_plot} (reproduced from Ref. \onlinecite{ritz2020strain}) shows that for a tensile misfit strain of 1\% at 300 K, corresponding approximately to growth of PbTiO$_3$ on the substrate DyScO$_3$, $T_c$ can be either higher or lower than bulk PbTiO$_3$, depending on the value of $\alpha_a$.

There is an intimate and complex connection between the transition temperature, the misfit strain, the thermal expansion coefficient of the substrate, and the thermal expansion coefficient of the out-of-plane $c$-axis ($\alpha_c$) of the PbTiO$_3$ thin-film. Rather than continuing to focus on transition temperatures, in this work we shift our attention to the fundamental relationship between the misfit strain, $\alpha_c$ and the thermal expansion coefficient of the substrate. Our goal is to elucidate the microscopic origins of this relationship in terms of the elastic and vibrational properties of PbTiO$_3$ thin-films. That is, we aim to answer the question, how can we understand differences in the thermal expanion properties of PbTiO$_3$ thin-films compared to bulk in terms of fundamental elasticity and vibrational properties?

\subsection{The thermal expansion and Gr\"{u}neisen tensors}
We begin by describing the thermodynamic framework through which we will explore and answer the question posed above. In the Gr\"{u}neisen theory of thermal expansion \cite{wallace,horton1974dynamical}, the thermal expansion can be related to a thermal stress proportional to the mean (or bulk) Gr\"uneisen parameters. The generalized mode Gr\"uneisen parameter \cite{wallace} for phonon mode $s$ at wave vector $\mathbf{q}$ is defined as the change in phonon frequency $\omega$ with respect to an infinitesimal strain $\epsilon$ along Cartesian directions $i$ and $j$:
\begin{equation}
\gamma^{ij}_{s,\mathbf{q}}\equiv-\frac{1}{\omega_{s,\mathbf{q}}}\frac{\partial \omega_{s,\mathbf{q}}}{\partial\varepsilon_{ij}}.    
\end{equation}
The mean Gr\"{u}neisen parameter \cite{choy1984thermal,ashcroft2005solid} is then essentially the sum of the individual mode Gr\"{u}neisen parameters weighted by their specific heats, $c_{s,\mathbf{q}}$:
\begin{equation}
    \gamma^{ij}\equiv \frac{\sum_{s,\mathbf{q}}\gamma^{ij}_{s,\mathbf{q}}c_{s,\mathbf{q}}}{\sum_{s,\mathbf{q}}c_{s,\mathbf{q}}}.
    \label{bulk_g}
\end{equation}
At a given temperature, the mean Gr\"{u}neisen tensor can be related to the thermal strain tensor, $\alpha_{j}$, through the elastic stiffness tensor $C_{ij}$,\cite{wallace,munn1972role}

\begin{equation} \sum_j C_{ij}\,\alpha_j=\frac{C_{\eta}}{V}\gamma^i,
\label{eq:GrunOrig}
\end{equation}
where $C_\eta$ is the bulk heat capacity at constant configuration, $V$ is the equilibrium volume and $\gamma^i$ is the mean Gr\"{u}neisen parameter of Equation \ref{bulk_g} in Voigt notation. Both $\alpha_{j}$ and $\gamma^i$ are 6$\times$1 vectors and $C_{ij}$ is a 6$\times$6 tensor, again employing Voigt notation. We emphasize that all of the materials properties in Equation \ref{eq:GrunOrig} are temperature dependent, including the elastic constants.

Since thermal strains are forbidden by symmetry to induce shear strains in tetragonal systems, for this work we only need to consider the upper left 3$\times$3 block of the thermal expansion tensor. Hence, the thermal expansion tensor can be written as a 3$\times$1 vector where the first two indices correspond to the strain per degree Kelvin of the in-plane $a$ and $b$ lattice parameters (necessarily equal in a tetragonal system), and the third index corresponds to strain per degree Kelvin of the out-of-plane $c$ lattice parameter:
\begin{equation}
\alpha_j = \begin{bmatrix}
\alpha_a \\
\alpha_a \\
\alpha_c
\end{bmatrix}.
\end{equation}
The mean Gr\"uneisen tensor can similarly be represented as a 3$\times$1 vector where the first two indices correspond to the mean Gr\"uneisen parameter with respect to $a$ and $b$ (again, necessarily equal by symmetry) and the third index corresponds to the mean Gr\"uneisen parameter with respect to $c$:
\begin{equation}
   \gamma^i= \begin{bmatrix} 
\gamma^a \\
\gamma^a \\
\gamma^c
\end{bmatrix}.
\end{equation}
These relations help us understand how changes in the mechanical boundary conditions applied to PbTiO$_3$ affect its vibrational properties, and in this study allow us to quantify changes in the thermal expansion properties of the PbTiO$_3$ film as a function of misfit strain and substrate thermal expansion coefficient. 

We compute the materials properties discussed above using the results from the QHA. For $\alpha_j$, we compute a numerical derivative of strain at each value of $T$ using the predicted lattice parameters and a central finite difference method. The $C_{ij}$ relevant to the Gr\"uneisen framework of thermal expansion are calculated using derivatives of the Helmholtz free energy surface fit to the points in the quasiharmonic grid with respect to strain along the $a$ and $c$ axes (see Supplementary Information for details). Note that since the films are under a nonzero state of in-plane strain, the elastic stiffness tensor is no longer symmetric, \textit{i.e.} $C_{13} \neq C_{31}$, and requires both second derivatives of energy with respect to strain as well as first derivatives (stress) to calculate \cite{barron1965second,wallace1965lattice,wang2012nonlinear}.

After the $\alpha_j$ and $C_{ij}$ are calculated, the Gr\"uneisen parameters can be found by solving the system of linear equations described by Equation \ref{eq:GrunOrig}. In this study, we are often only concerned with analyzing thermal expansion along $c$, as $\alpha_a$ is defined by the substrate. Thus, the only Gr\"uneisen parameter necessary is $\gamma_c$, which can be solved for using the following relation derived from \ref{eq:GrunOrig}:
\begin{equation}
2C_{31}\alpha_a+C_{33}\alpha_c = \frac{C_{\eta}}{V}\gamma^c.
\end{equation}

\subsection{Relationship between linear elastic and anharmonic effects in the Gr\"{u}neisen theory}
Having established some basic definitions, what we would now like is an expression that relates changes (compared to bulk) in the thermal expansion properties of PbTiO$_3$ thin-films to corresponding changes in the elastic constants and vibrational properties (Gr\"{u}neisen parameters). Let $\Delta\alpha_i(T) \equiv \alpha_i(T)-\alpha_i^{\mathrm{bulk}}(T)$ such that $\alpha_i^{\mathrm{bulk}}(T)+\Delta\alpha_i(T) =\alpha_i(T)$. Similarly, let $\Delta C_{ij}(T) \equiv C_{ij}(T)-C_{ij}^{\mathrm{bulk}}(T)$, where C$_{ij}(T)$ is the temperature-dependent elastic stiffness tensor of the PbTiO$_3$ thin-film and $C_{ij}^{\mathrm{bulk}}(T)$ is the elastic stiffness tensor of bulk PbTiO$_3$. Finally, let $\Delta\gamma_i(T) \equiv \gamma_i(T)-\gamma_i^{\mathrm{bulk}}(T)$, where $\gamma_i(T)$ is the $i^{th}$ element of the Gr\"uneisen tensor for a PbTiO$_3$ thin-film and $\gamma_i^{\mathrm{bulk}}(T)$ is the Gr\"{u}neisen tensor of bulk PbTiO$_3$. Thus, $\Delta \alpha_i(T)$, $\Delta C_{ij}(T)$, and $\Delta \gamma_i(T)$ respectively denote the difference in thermal expansion, elastic properties, and Gr\"uneisen parameters in the strained film from that of bulk PbTiO$_3$ at the same temperature. 

In the Gr\"uneisen theory of thermal expansion, the difference in thermal expansion between a PbTiO$_3$ film and bulk PbTiO$_3$, $\Delta\alpha(T)$, is entirely accounted for by the departure of the Gr\"{u}nesien parameters and elastic constants from the values they would have in the bulk system (with higher-order phonon-phonon coupling terms ignored \cite{wallace, horton1974dynamical}). Thus, using Equation \ref{eq:GrunOrig}, for the strained film
 
\begin{equation} \sum_j C_{ij}(T)\,\alpha_j(T)=\frac{C_{\eta}(T)}{V(T)}\gamma_i(T),
\end{equation}
and therefore,
\begin{multline} \sum_j (C_{ij}^{\mathrm{bulk}}(T)+\Delta C_{ij}(T))\,(\alpha^{\mathrm{bulk}}_j(T)+\Delta\alpha_j(T))= \\
\frac{C_{\eta}(T)}{V(T)}(\gamma^{\mathrm{bulk}}_i(T)+\Delta \gamma_i(T)).
\label{eq:GrunPrimed}
\end{multline}
Expanding Equation \ref{eq:GrunPrimed} for $i=3$ for a tetragonal system and subtracting from it the relation in Equation \ref{eq:GrunOrig} for the bulk system, 

\begin{align}  
\begin{split}
2C^{\mathrm{bulk}}_{31}\Delta \alpha_a + C^{\mathrm{bulk}}_{33}\Delta \alpha_c+
2\Delta C_{31}\alpha^{\mathrm{bulk}}_a + \Delta C_{33}\alpha^{\mathrm{bulk}}_c+\\
2\Delta C_{31} \Delta \alpha_a + \Delta C_{33}\Delta \alpha_c = \frac{C_{\eta}}{V}\Delta\gamma^c.
\label{eq:delta_Cgamma}
\end{split}
\end{align}
\noindent
We remind the reader again that since the film is under non-zero stress along the $a$-axis, the elasticity tensor is no longer necessarily symmetric, and thus $C_{31} \neq  C_{13}$. For notational simplicity, we have also dropped the $(T)$ from each of these quantities, though they all remain functions of temperature. Solving Equation \ref{eq:delta_Cgamma} for $\Delta \alpha_c$ and rearranging (see Supplementary Information for a more detailed derivation of Equations \ref{eq:GrunPrimed}--\ref{eq:full_exp}), we obtain,

\begin{equation}  
\begin{split}  
 \Delta \alpha_c= -2\frac{C^{\mathrm{bulk}}_{31}}{C^{\mathrm{bulk}}_{33}}\Delta \alpha_a+\\\left[\left(2\frac{C^{\mathrm{bulk}}_{31}}{C^{\mathrm{bulk}}_{33}}\Delta \alpha_a-\alpha^{\mathrm{bulk}}_c\right)\frac{\Delta C_{33}}{(C^{\mathrm{bulk}}_{33}+\Delta C_{33})}\right. \\\left. -2(\alpha^{\mathrm{bulk}}_a + \Delta \alpha_a)\frac{\Delta C_{31} }{(C^{\mathrm{bulk}}_{33}+\Delta C_{33})}  \right]+\\     \frac{C_{\eta}\Delta\gamma^c}{V (C^{\mathrm{bulk}}_{33}+\Delta C_{33})}.
\end{split}
\label{eq:full_exp}
\end{equation}

\noindent
 Though it appears complex, Equation \ref{eq:full_exp} establishes the sought after connection between $\Delta \alpha_c$, the \emph{deviation} of $\alpha_c$ of the strained film from that of bulk PbTiO$_3$, and the elastic and vibrational properties of strained PbTiO$_3$ films and how \emph{they} depart from that of bulk PbTiO$_3$. By understanding the physical meaning of each term in Equation \ref{eq:full_exp}, we can explain the microscopic mechanism underlying $\Delta \alpha_c$ in strained film systems in terms of changes in elasticity and vibrational properties:

\begin{enumerate}

    \item We refer to the first term, $-2\frac{C^{\mathrm{bulk}}_{31}}{C^{\mathrm{bulk}}_{33}}\Delta \alpha_a$, as \emph{linear elastic} coupling. This is the elastic coupling between in-plane ($a$-axis) and out-of-plane ($c$-axis) thermal strain from linear elasticity. It accounts for the change in the rate of thermal expansion along $c$ of the film that would be predicted if the materials properties $C_{ij}$ and $\gamma^c$ did not change at all with strain.

    \item We refer to the second and third terms (grouped by square brackets in Equation \ref{eq:full_exp}) as \emph{elastic anharmonicity}. These terms capture how changes in the elastic properties in the strained film affect $\Delta \alpha_c$.
    
    \item We refer to the last term, $-\frac{C_{\eta}\Delta\gamma^c}{V (C^{\mathrm{bulk}}_{33}+\Delta C_{33})}$, as \emph{anharmonic thermal stress}. This term captures how changes in phonon anharmonicity (due to phonon-strain coupling), and the thermal stress it induces in the strained film, affect $\Delta \alpha_c$.
\end{enumerate}

It is clear that the difference in the $c$-axis thermal expansion of strained PbTiO$_3$ thin-films compared to bulk arises from three sources -- the mismatch between the thermal expansion coefficients of the substrate and bulk PbTiO$_3$ ($\Delta \alpha_a$), the difference between the elastic constants of the strained film compared to those of bulk PbTiO$_3$ ($\Delta C_{31}$ and $\Delta C_{33}$), and the difference between the Gr\"uneisen parameters along the $c$-axis of the strained film compared to those of bulk PbTiO$_3$ ($\Delta \gamma_{c}$). Note that the misfit strain at a given temperature gives rise to changes in the elastic constants and the Gr\"{u}neisen parameters ($\Delta \gamma^c$, $\Delta C_{31}$ and $\Delta C_{33}$) at that temperature, whereas $\Delta \alpha_a$ determines the misfit strain \textit{rate}. In the following section, we consider how each of $\Delta C_{31}$, $\Delta C_{33}$ $\Delta \gamma^{c}$, and $\Delta \alpha_a$ affect the thermal expansion of strained PbTiO$_3$ thin-films along the $c$-axis, and how they depend on misfit strain and the thermal expansion coefficient of a given substrate.

\section{Results}
\subsection{Thermal expansion in strained thin-films -- Importance of anharmonicity}
\label{ssec:understanding}
\begin{figure}
    \includegraphics[width=9cm]{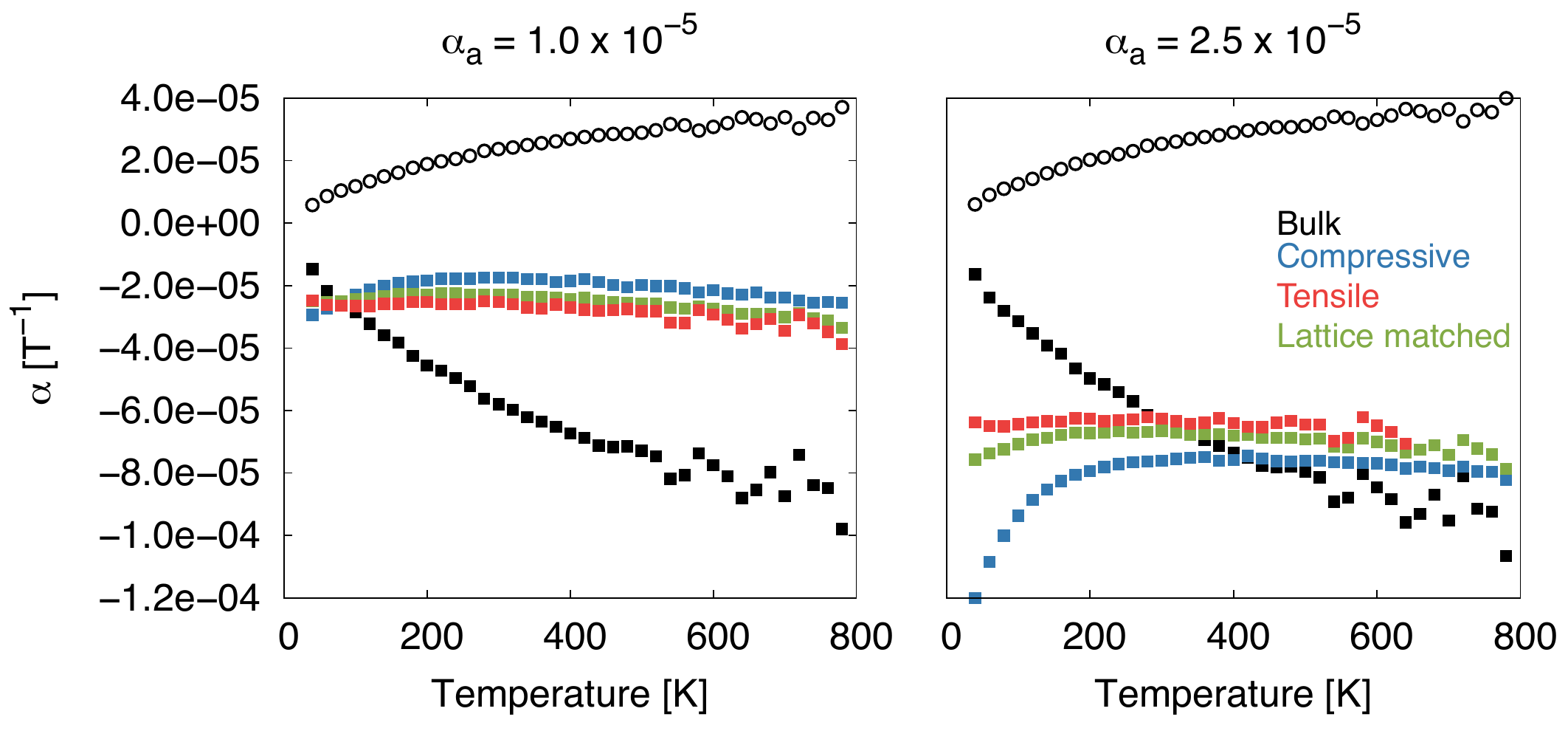}
    \caption{Comparison between $c$-axis thermal expansion coefficients ($\alpha_c$) as a function of temperature for bulk PbTiO$_3$ and a series of strained films from our QHA calculations for growth on two different substrates with $\alpha_a$ = 1.0$\times$10$^{-5}$ (left) and 2.5$\times$10$^{-5}$ (right) -- compressive ($\varepsilon_a^{300}$ = -0.75\%, tensile ($\varepsilon_a^{300}$ = +0.75\%) and lattice-matched ($\varepsilon_a^{300}$ = 0.00\%). The $\alpha_c$ data are denoted by closed squares, whereas the $\alpha_a$ data (shown for bulk PbTiO$_3$ only) are denoted by open black circles. The data for bulk PbTiO$_3$ are identical in both panels.}
    \label{fictitious_lats}
\end{figure}

Figure \ref{fictitious_lats} shows the rate of thermal expansion along the $c$-axis for PbTiO$_3$ thin-films as a function of strain and two different substrate thermal expansions compared with bulk PbTiO$_3$. When $\alpha_a$ = 1$\times$10$^{-5}$, the negative thermal expansion along $c$ is suppressed compared to bulk for each of the strained films. When the rate of thermal expansion of the substrate is increased ($\alpha_a$ = 2.5$\times$10$^{-5}$), the strained films initially show an enhanced rate of negative thermal expansion along $c$ compared to bulk, up to about 400 K. Above this temperature, bulk PbTiO$_3$ once again shows the largest rate of negative thermal expansion along $c$. The behavior of the lattice-matched films (zero misfit strain at 300 K) is always `between' that of the tensile and compressively strained films. In order to prevent the discussion below from becoming overly complex, we do not explicitly discuss the lattice-matched films.\cite{strain}

\begin{figure}
    \includegraphics[width=9cm]{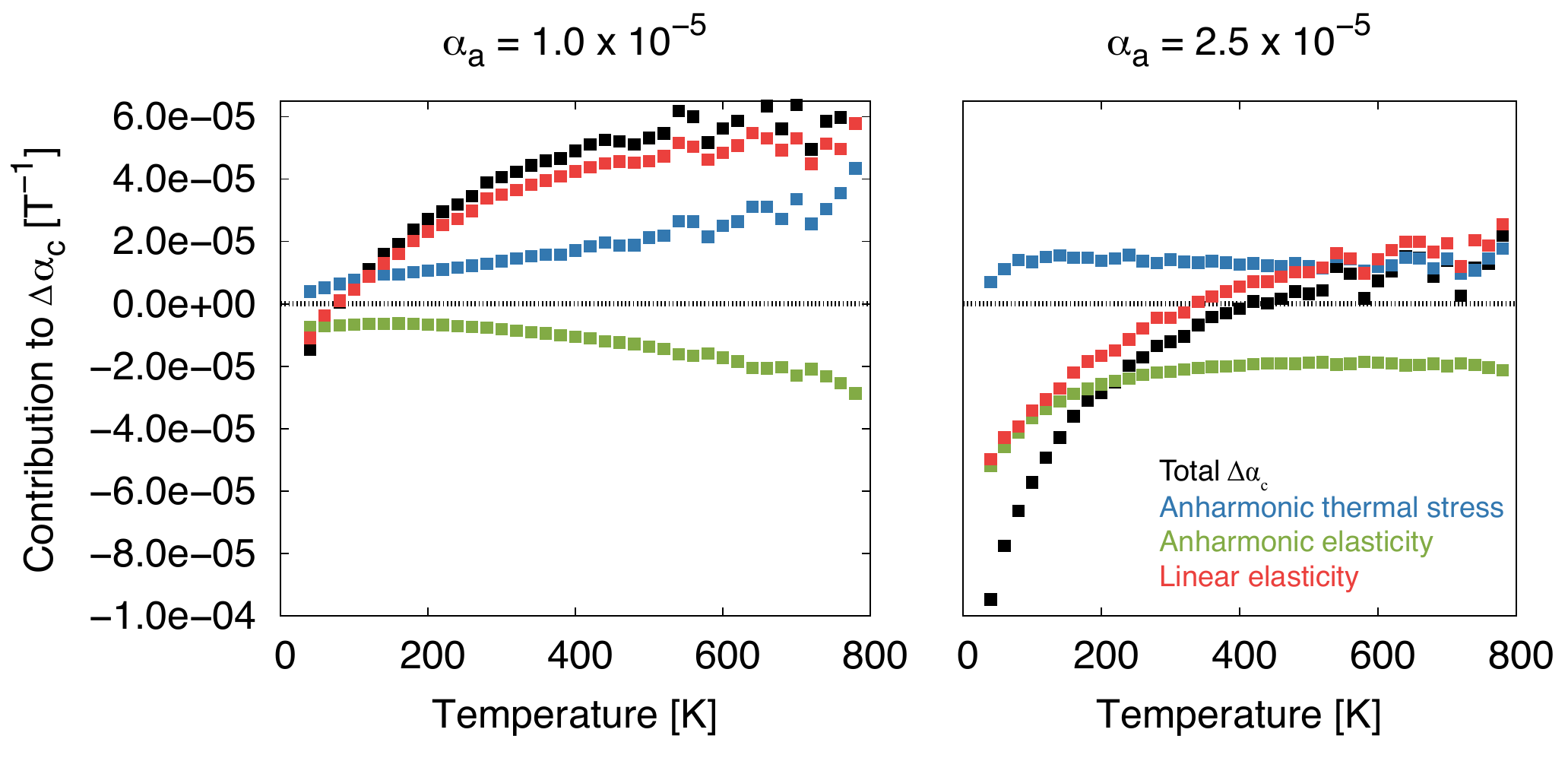}
    \caption{Contributions to $\Delta \alpha_c$, broken down according to Equation \ref{eq:full_exp}, as a function of temperature for compressively strained films. The misfit strain at 300 K ($\varepsilon_a^{300}$) is -0.75\% while the strain rate is varied from that of bulk PbTiO$_3$. The linear elasticity contribution refers to the first term of Equation \ref{eq:full_exp}, the anharmonic elasticity contribution refers to the second and third terms of Equation \ref{eq:full_exp} (grouped in parentheses), and the anharmonic thermal stress contribution refers to the fourth term of Equation \ref{eq:full_exp}.}.
    \label{vary_eps0_m}
\end{figure}

\begin{figure}
    \includegraphics[width=9cm]{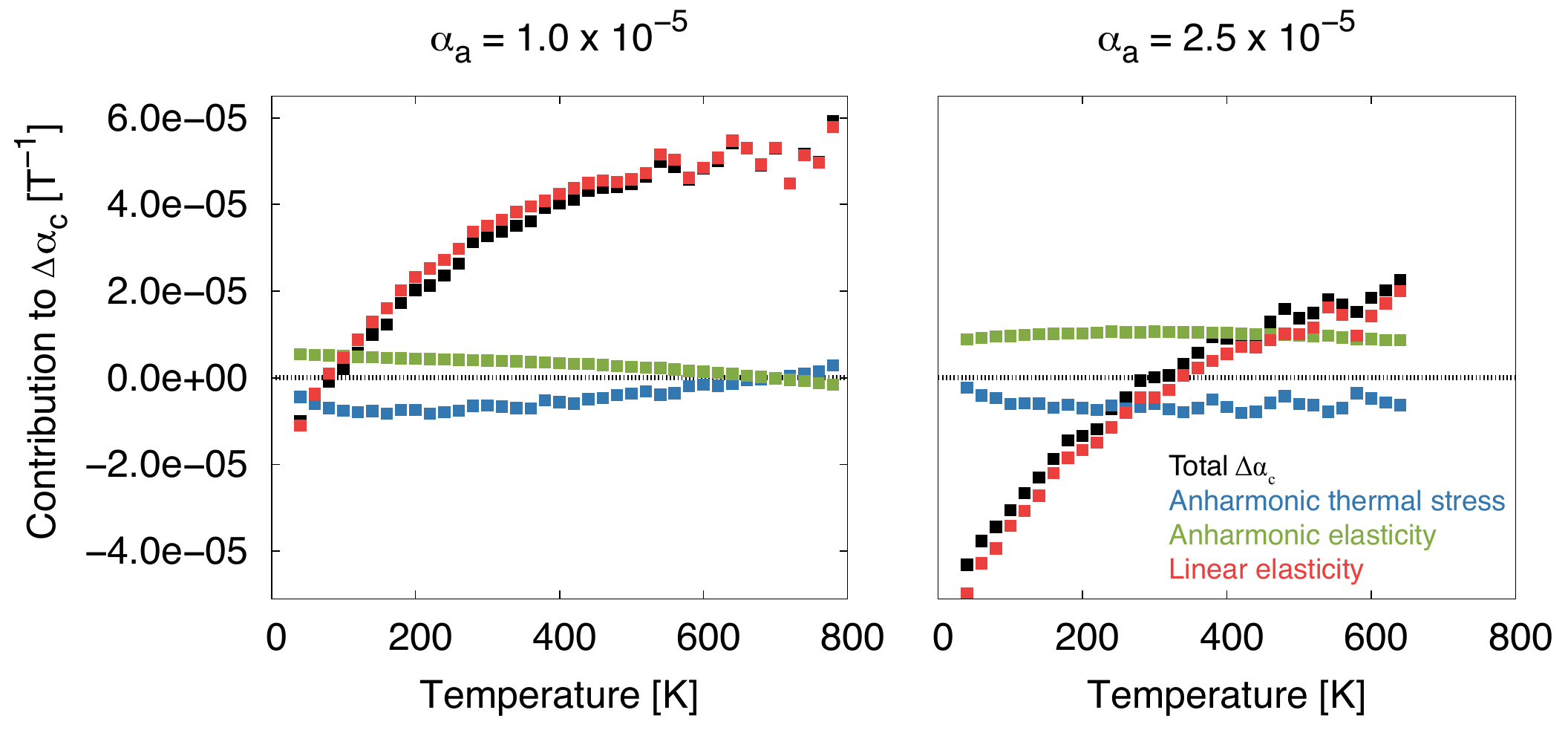}
    \caption{Contributions to $\Delta \alpha_c$, broken down according to Equation \ref{eq:full_exp}, as a function of temperature for tensile strained films. The misfit strain at 300 K ($\varepsilon_a^{300}$) is +0.75\%, while the strain rate is varied from that of bulk PbTiO$_3$. The linear elasticity contribution refers to the first term of Equation \ref{eq:full_exp}, the anharmonic elasticity contribution refers to the second and third terms of Equation \ref{eq:full_exp} (grouped in parentheses), and the anharmonic thermal stress contribution refers to the fourth term of Equation \ref{eq:full_exp}.}.
    \label{vary_eps0_p}
\end{figure}

We can understand the trends shown in Figure \ref{fictitious_lats} by considering individually the contributions of elastic anharmonicity, thermal stress and linear elasticity in Equation \ref{eq:full_exp} to $\Delta\alpha_c$, the deviation in thermal expansion along $c$ in strained PbTiO$_3$ compared to bulk. Figures \ref{vary_eps0_m} and \ref{vary_eps0_p} show $\Delta\alpha_c$ and its various contributions for PbTiO$_3$ thin-films under compressive and tensile strain for two different substrate thermal expansion coefficients (the same as those considered in Figure \ref{fictitious_lats}). Both figures appear to show that linear elasticity (the first term of Equation \ref{eq:full_exp}) makes the dominant contribution to $\Delta\alpha_c$, with the red squares almost overlaid on the black squares. In this scenario, the thermal expansion of the $c$-axis as a function of temperature is controlled by Poisson-like linear elastic coupling between the $a$ and $c$-axes. If we were to consider \emph{only} the linear elastic contribution to $\Delta\alpha_c$, the PbTiO$_3$ films would appear highly harmonic. 

However, Figures \ref{vary_eps0_m} and \ref{vary_eps0_p} additionally show that there are also large contributions to $\Delta\alpha_c$ from elastic anharmonicity and anharmonic thermal stress -- in the case of the PbTiO$_3$ thin-film under compressive strain and $\alpha_a$ = 2.5$\times$10$^{-5}$, these terms are actually larger in magnitude than the linear elastic contribution through nearly the entire temperature range. Critically, the contributions to $\Delta\alpha_c$ from elastic anharmonicity and anharmonic thermal stress are of \emph{opposite sign}, leading to a fortuitous near-cancellation (whereas the sign of the contribution from elastic anharmonicity is always opposite that of the anharmonic thermal stress, which one is positive and which negative depends on the sign of the misfit strain at a given temperature; for elastic anharmonicity the contribution tends to be the same sign as the misfit strain). This fortuitous near-cancellation between different types of anharmonicity, the origin of which will be discussed in the next section, results in what looks like a linear elastic thermal response of strained PbTiO$_3$ thin-films.

\subsection{Dependence of elastic constants and Gr\"{u}neisen parameters on temperature and misfit strain}\label{ssec:dependence}
\begin{figure*}
    \includegraphics[width=\textwidth]{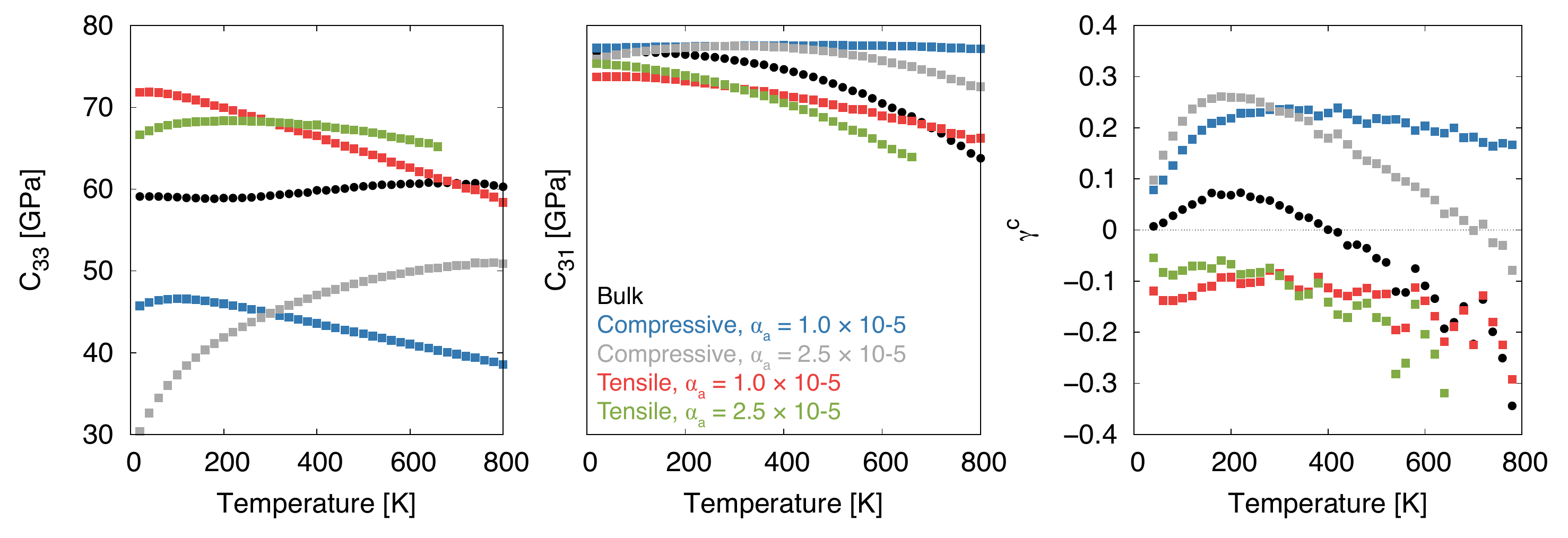}
    \caption{Comparison between strained films on substrates with two different thermal expansion coefficients ($\alpha_a = 1.0\times10^{-5}$ and $\alpha_a=2.5\times10^{-5}$) and bulk PbTiO$_3$ for $C_{33}$ (left), $C_{31}$ (middle) and $\gamma_c$ (right) from our first-principles calculations.}
    \label{fig:change_elast_gamma}
\end{figure*}

The near-cancellation of contributions from anharmonicity shown in Figures \ref{vary_eps0_m} and \ref{vary_eps0_p} can ultimately be attributed to changes in the elastic constants and Gr\"{u}neisen parameters with temperature and misfit strain. Figure \ref{fig:change_elast_gamma} shows that $C_{31}$, $C_{33}$ and $\gamma_c$ for the strained films can all differ significantly from bulk PbTiO$_3$, and also change significantly with temperature depending on the misft strain and substrate thermal expansion coefficient. Despite the fact that contributions to $\Delta\alpha_c$ from just the linear elasticity term of Equation \ref{eq:full_exp} closely match those calculated using the full expression for $\Delta\alpha_c$, the large strain dependencies depicted in Figure \ref{fig:change_elast_gamma} corroborate the data shown in Figures \ref{vary_eps0_m} and \ref{vary_eps0_p} in illustrating that strained PbTiO$_3$ thin-films are highly anharmonic, especially with respect to the elastic constants (note that even though the values of $\gamma_c$ tend to be fairly small, they still experience large \emph{changes} with strain, indicating the presence of anharmonicity). The details of how the elastic constants and Gr\"{u}neisen parameters change with temperature affect how elastic anharmonicity and anharmonic thermal stress contribute to changes in the $c$-axis of strained PbTiO$_3$ films with temperature.

 Considering first the compressively strained film, $C_{33}$ is softer than in bulk PbTiO$_3$ ($\Delta C_{33} < 0$), whereas $C_{31}$ is stiffer ($\Delta C_{31} > 0$). The effect of these strain-induced changes to elasticity, which is captured by the second term (grouped in square brackets) in Equation \ref{eq:full_exp} and the green squares in Figure \ref{vary_eps0_m}, is to make a negative contribution to $\Delta\alpha_c$. Films under tensile strain exhibit the opposite behavior of the elastic constants compared to bulk -- under tensile strain, elastic anharmonicity makes a positive contribution to $\Delta\alpha_c$, as shown by the green squares in Figure \ref{vary_eps0_p}. The Supplementary Information contains a short discussion of how the observed effect of changes in the elastic constants on $\Delta\alpha_c$ can be predicted from a careful analysis of the sign of each of the two terms contributing to anharmonic elasticity in Equation \ref{eq:full_exp}.

With respect to the Gr\"{u}neisen parameters, Figure \ref{fig:change_elast_gamma} shows that $\gamma_c$ is larger in the compressively strained film than in bulk PbTiO$_3$ ($\Delta\gamma_c > 0$), whereas it is smaller in the tensile strained film ($\Delta\gamma_c < 0$). Referring back to Equation \ref{eq:full_exp}, compressive strain then causes the anharmonic thermal stress (last) term of Equation \ref{eq:full_exp} to make a positive contribution to $\Delta\alpha_c$, as shown by the blue squares in Figure \ref{vary_eps0_m}. In contrast, the anharmonic thermal stress makes a negative contribution to $\Delta\alpha_c$ in films under tensile strain, as shown by the blue squares in Figure \ref{vary_eps0_p} (note that while $C_{33}$ appears in the denominator of this term, it cannot change the sign of the term). In other words, the contribution from anharmonic thermal stress to thermal expansion along $c$ will be the same sign as $\Delta \gamma_c$, which is typically opposite in sign to the misfit strain. The changes in contributions from the anharmonic thermal stress with respect to misfit strain are opposite those of anharmonic elasticity, as discussed above.

Hence, when PbTiO$_3$ undergoes biaxial strain, there is a competition between elastic anharmonicity (the second and third terms of Equation \ref{eq:full_exp}), and anharmonic thermal stress (the last term of Equation \ref{eq:full_exp}),  and a large portion of their sum cancels. This near-cancellation is the result of large, simultaneous changes in elastic and vibrational properties in the film, and causes the response to appear nearly elastic, with a small departure from linear elasticity due to the cancellation not being exact. The contribution to $\alpha_c$ from anharmonic elasticity is proportional to $\Delta C_{33}$, which is typically the same sign as the misfit strain, and opposes the contribution from anharmonic thermal stress proportional to $\Delta \gamma_c$, which is typically the opposite sign of the misfit strain. 

\subsection{Understanding differences among strained films in terms of contributions to $\Delta\alpha_c$}

Finally, returning now to Figure \ref{fictitious_lats}, although all of the strained films we studied exhibit suppressed negative thermal expansion along the $c$-axis compared to bulk (except at low temperatures for the $\alpha_a$ = 2.5$\times$10$^{-5}$ substrate), there are differences in thermal expansion among the strained films, as noted earlier. If the behavior of the PbTiO$_3$ thin-films really was fully controlled by linear elasticity, the thermal expansion of the $c$-axis would be completely determined by the substrate thermal expansion coefficient $\alpha_a$ and the elastic properties of bulk PbTiO$_3$. Hence, the $c$-axis thermal expansion would be the same, regardless of the misfit strain. Instead, the details of the thermal expansion of the films do depend on both the misfit strain and the mismatch between the thermal expansion coefficients of the substrate and bulk PbTiO$_3$ ($\Delta\alpha_a$, the misfit strain \emph{rate}). 

For example, compressively strained films on substrates with $\alpha_a$~=~ 1$\times$10$^{-5}$ show the \emph{least} negative rate of thermal expansion along the $c$-axis but when $\alpha_a$ increases to 2.5$\times$10$^{-5}$, they show the \emph{most} negative thermal expansion along $c$ among the thin-film systems. This is because for compressive strain and $\alpha_a$~=~ 1$\times$10$^{-5}$, the positive contributions to $\Delta\alpha_c$ from anharmonic thermal stress are larger in magnitude than the negative contributions from anharmonic elasticity, pushing the thermal expansion of the $c$-axis to be slightly more positive than for the other strained systems, as shown in the left panel of Figure \ref{vary_eps0_m}. When $\alpha_a$~=~ 2.5$\times$10$^{-5}$, the negative contributions to $\Delta\alpha_c$ from elastic anharmonicity are larger in magnitude than the positive contributions from anharmonic thermal stress, pushing the thermal expansion of $c$ to be slightly more negative than for the other strained systems, as shown in the right panel of Figure \ref{vary_eps0_m}. These trends are reversed for films under tensile strain.

\begin{figure}
    \includegraphics[width=9cm]{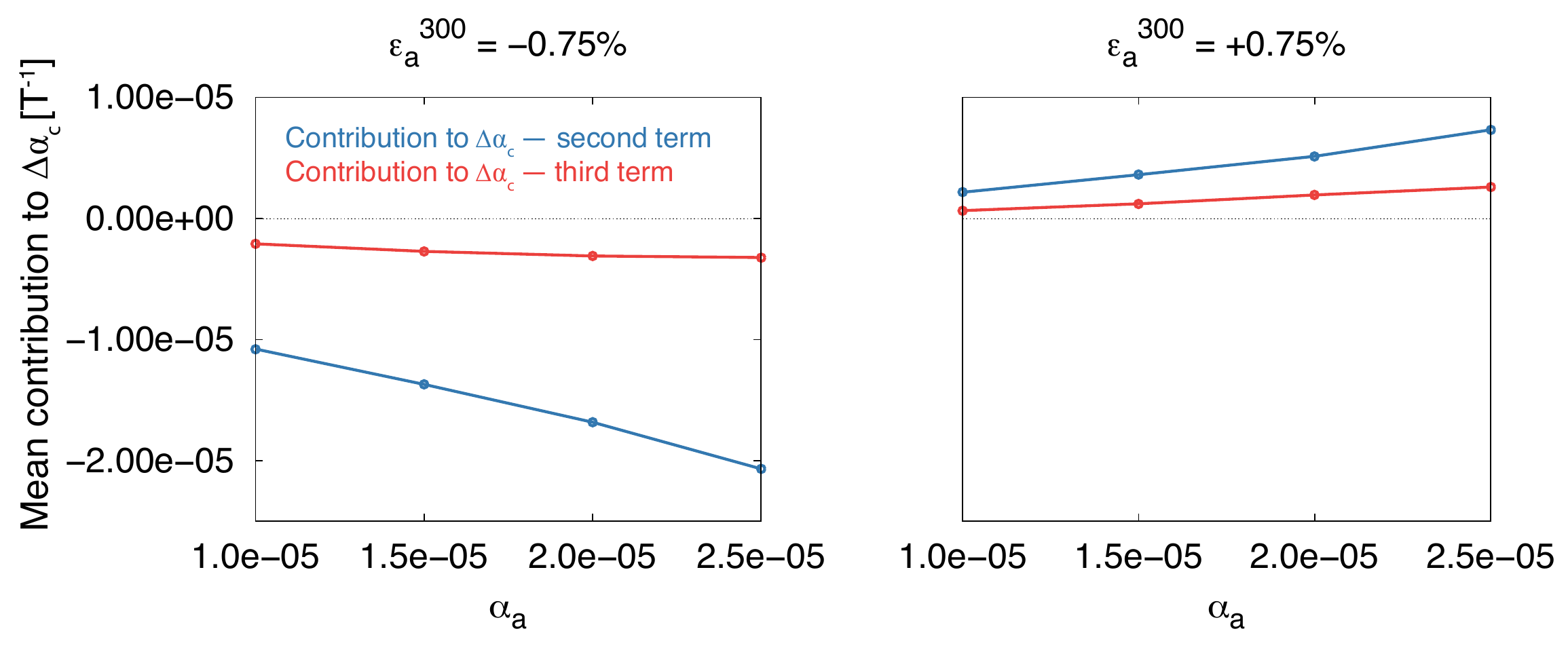}
    \caption{Variation in contribution to $\Delta\alpha_c$ of anharmonic elasticty term of Equation \ref{eq:full_exp} in terms of the second term of Equation \ref{eq:full_exp} (blue) and the third term of Equation \ref{eq:full_exp} (red). Data are shown as a function of substrate thermal expansion coefficient ($\alpha_a$) for compressively strained ($\epsilon_a^{300}$ = -0.75\%) and tensile strained ($\epsilon_a^{300}$ = +0.75\%) films. Each data point corresponds to the average value of the indicated term from 0 K to 800 K. The lines are guides for the eye.}
    \label{fig:mean_cont_elas}
\end{figure}

Figures \ref{vary_eps0_m} and \ref{vary_eps0_p} show that, for a given strain (compressive or tensile), the anharmonic thermal stress is relatively insensitive to $\alpha_a$. In addition, Equation \ref{eq:full_exp} shows that neither $\alpha_a$ nor $\Delta\alpha_a$ appear in the (fourth) anharmonic stress term. This directs our attention to the two terms associated with anharmonic elasticity. Figure \ref{fig:mean_cont_elas} shows how the average values of the two contributions to $\Delta\alpha_c$ from anharmonic elasticity terms in Equation \ref{eq:full_exp} change as a function of $\alpha_a$ for both compressive and tensile strain.  For films under compressive strain, it is the first term, $ \left(2\frac{C^{\mathrm{bulk}}_{31}}{C^{\mathrm{bulk}}_{33}}\Delta \alpha_a-\alpha^{\mathrm{bulk}}_c\right) \Delta C_{33}/(C^{\mathrm{bulk}}_{33}+\Delta C_{33})$, that has the most negative contribution to $\Delta\alpha_c$ and the most negative slope with respect to $\alpha_a$. This means that for compressively strained films, it is this term that makes the contribution of elastic anharmonicity to $\Delta\alpha_c$ negative and larger in magnitude than the positive contribution from anharmonic thermal stress. Although the compressively strained film initially shows the least negative rate of thermal expansion along the $c$-axis among the strained films, as $\alpha_a$ increases the magnitude of the negative contribution from elastic anharmonicity also increases, leading the compressively strained film to exhibit the most negative rate of thermal expansion along $c$ when $\alpha_a$ = 2.5$\times$10$^{-5}$. For films under tensile strain, both anharmonic elastic terms are positive and increase as $\alpha_a$ increases -- the first term again exhibits the largest contribution and slope with respect to $\alpha_a$, yet this time both are positive. Hence, in this case the positive contributions to $\Delta\alpha_c$ from anharmonic elasticity grow larger in magnitude than the negative contributions from anharmonic thermal stress as $\alpha_a$ increases, leading to films under tensile strain having a slower rate of thermal expansion along the $c$-axis than either bulk PbTiO$_3$ or compressively strained films. As discussed in Section \ref{ssec:dependence} and the Supplementary Information, the sign of this critical first term is determined by the sign of $\Delta C_{33}$, which is the same as the sign of $\varepsilon_a$. The behavior of this term, combined with the relative insensitivity of anharmonic thermal stress on $\alpha_a$, explains the departure of $\Delta\alpha_c$ from what would be predicted by linear elasticity alone.

\subsection{Explaining thermal expansion in experimentally synthesized PbTiO$_3$ thin-films}\label{ssec:experimental}
\begin{figure}
    \includegraphics[width=8.5cm]{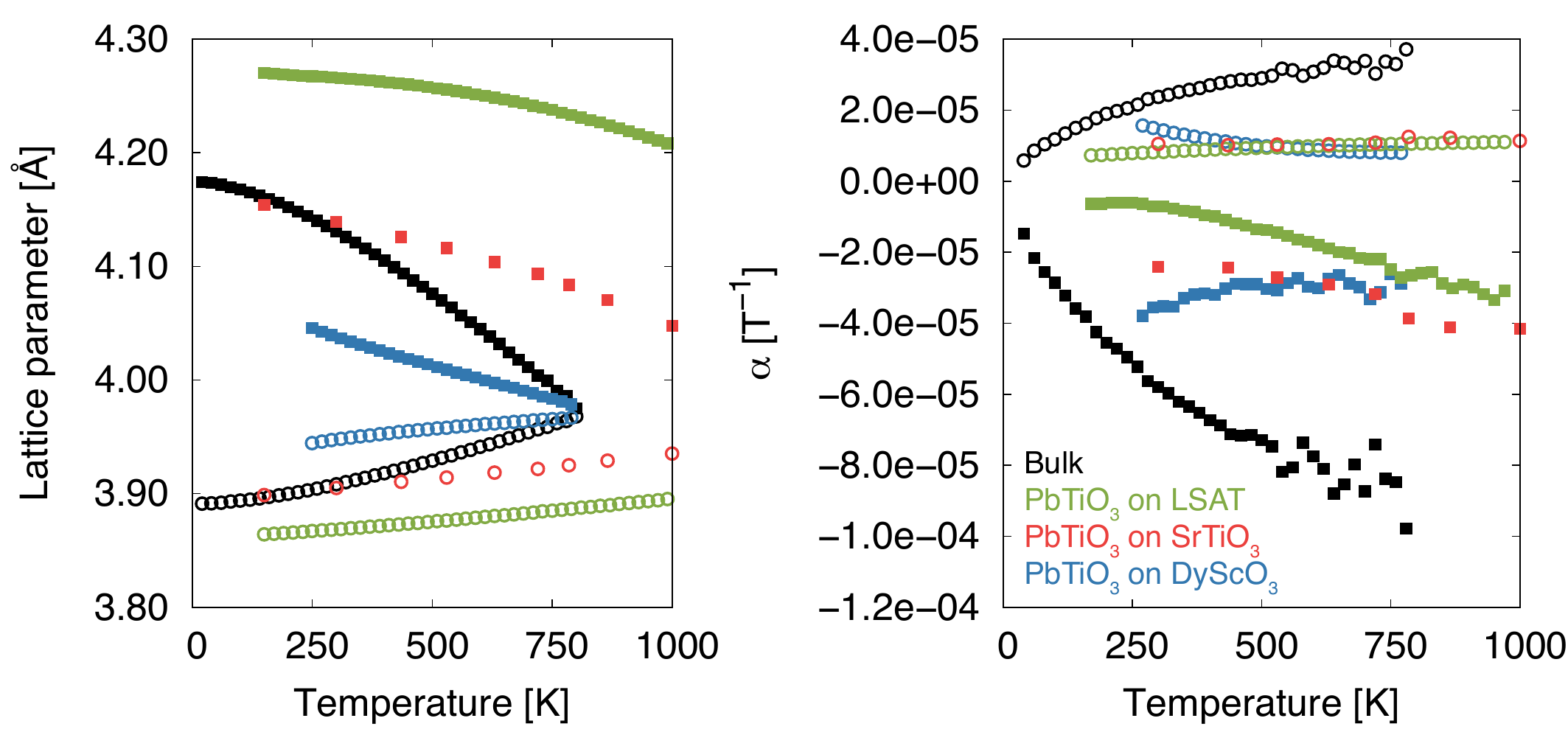}
    \caption{Comparison between lattice parameters (left) and thermal expansion coefficients (right) of bulk PbTiO$_3$ and strained PbTiO$_3$ thin-films on LSAT, DyScO$_3$ and SrTiO$_3$ substrates. Substantially similar data are reported in Ref. \onlinecite{ritz2020strain}, however in this work we used a slightly more dense grid for our QHA calculations. Open cirlces denote data for the $a$-axis lattice parameters and $\alpha_a$ (experimental data for SrTiO$_3$, DyScO$_3$, and LSAT\cite{de1996high,biegalski2005thermal,chakoumakos1998thermal}), whereas closed squares denote data for the $c$-axis lattice parameters and $\alpha_c$ (predicted from our QHA calculations).}
    \label{alphas_realsubs}
\end{figure}

We can use our findings above to explain the experimentally observed thermal expansion behavior of PbTiO$_3$ thin-films on LSAT,\cite{chakoumakos1998thermal} SrTiO$_3$\cite{de1996high} and DyScO$_3$\cite{biegalski2005thermal} substrates. Figure \ref{alphas_realsubs} shows the lattice parameters and thermal expansion coefficients as functions of temperature for PbTiO$_3$ thin-films on all three substrates. All three substrates have similar thermal expansion coefficients -- from 0 -- 800 K, $\alpha_a$ is fairly low and constant with temperature with average values of $0.93\times10^{-5}$ K$^{-1}$, $1.10\times10^{-5}$ K$^{-1}$, and $1.04\times10^{-5}$ K$^{-1}$, for LSAT, SrTiO$_3$, and DyScO$_3$, respectively. In each case the thermal expansion along $c$ is less negative than in bulk PbTiO$_3$, however the $\alpha_c$ for each substrate is different, and the framework established in this study explains why.

At 300 K, LSAT induces a large compressive misfit strain of -1.01\%, SrTiO$_3$ a small strain of -0.01\%, and DyScO$_3$ induces a large tensile strain of +1.01\%. We showed above that under compressive strain and when $\alpha_a$ is small, the negative magnitude of the contribution to $\Delta\alpha_c$ from elastic anharmonicity is smaller than the positive contribution from anharmonic thermal stress, leading to a smaller rate of negative thermal expansion along $c$ compared to both bulk and tensile strained films. By this reasoning, PbTiO$_3$ on LSAT should show the least negative thermal expansion along $c$ and PbTiO$_3$ on DyScO$_3$ should show the most negative thermal expansion along $c$; this is indeed what is shown in Figure \ref{alphas_realsubs}. Put another way, we showed earlier that the sign of the contribution from anharmonic thermal stress to $\Delta \alpha_c$ has the opposite sign of the misfit strain. Therefore, we would expect $\Delta \alpha_c$ to be most positive in LSAT and most negative in DyScO$_3$, and this is indeed what we see.

\section{Summary and Conclusions}
Our results reveal that although PbTiO$_3$ thin-films are highly anharmonic, a fortuitous near-cancellation between the thermal stress and anharmonic elastic contributions to the temperature-dependent structural parameters means that the thermal response of the $c$-axis of the films can be predicted very well using only the misfit strain and bulk PbTiO$_3$ properties. We note that the $\Delta \gamma^c$ term, which contributes to the anharmonic thermal stress, involves the \emph{derivative} of the Gr\"uneisen parameter with respect to strain. Since the Gr\"uneisen parameter is a function of the third-order interatomic force constants (and higher order, if sublattice displacements are present, as is the case for ferroelectric PbTiO$_3$ \cite{horton1974dynamical,wallace}), $\Delta \gamma^c$ is a function of the \emph{fourth} and higher-order force constants. Additionally, as elastic constants are related to the curvature of the Helmholtz free energy surface, the large changes in the elastic constants with strain are also driven by changes in both electronic and vibrational free energy, and thus the third, fourth, and higher-order force constants \cite{dangic2018coupling}. The low thermal conductivity observed for bulk PbTiO$_3$\cite{tachibana2008thermal,yoshida1960thermal} corroborates these findings. In this light however, the approximately linear-elastic behavior of the strained PbTiO$_3$ thin-films is unexpected. 

Our findings may also help explain why popular phenomenological models for predicting the strain behavior of the polarization in ferroelectric PbTiO$_3$ thin-films, such as those described in Refs. \citenum{pertsev1998effect} and \citenum{waser02}, are reasonably effective at capturing the thermal expansion behavior of both bulk and epitaxially strained films of PbTiO$_3$ with temperature \cite{highland2014interfacial,lichtensteiger2005ferroelectricity,venkatesan2008monodomain,haun1987thermodynamic}, despite being functions of the elastic constants of the high-symmetry cubic phase, and with no explicit elastic anharmonicity included beyond its coupling to the polarization. If the structural behavior of ferroelectric PbTiO$_3$ films clamped to a substrate can be well-described by linear elasticity with much of the higher-order anharmonicity ignored, it stands to reason that the inclusion of higher-order elastic coupling terms beyond the few judiciously chosen in the above models would be unnecessary to accurately capture the structural evolution of the system with temperature.

Finally, our work shows that the thermal expansion behavior of PbTiO$_3$ films depends on the misfit strain throughout the \emph{entire} temperature range. This means it depends on not only the misfit strain between film and substrate lattice parameters at a single temperature, but on the misfit between their thermal strain \emph{rates}, which determines how misfit strain accumulates with temperature. These dependencies are entwined in a complex way -- the misfit strain between the lattice parameters of the substrate and the equilibrium (bulk) lattice parameters of the film material can induce large changes in the elastic properties and Gr\"uniesen parameters of the film, which themselves influence how the $c$-axis of the film responds to the misfit strain rate. Our combined first-principles and phenomenological approach provides significant insights into this behavior and offers a systematic framework for exploring thermal expansion phenomena in both bulk and thin-film systems.

\begin{acknowledgments}
This work was supported by the National Science Foundation. E. T. R. and N. A. B. were supported by DMR-1550347. Computational resources were provided by the Cornell Center for Advanced Computing and the Extreme Science and Engineering Discovery Environment (XSEDE) through allocation DMR-160052.
\end{acknowledgments}

\providecommand{\noopsort}[1]{}\providecommand{\singleletter}[1]{#1}%


\begin{thebibliography}{43}%
\makeatletter
\providecommand \@ifxundefined [1]{%
 \@ifx{#1\undefined}
}%
\providecommand \@ifnum [1]{%
 \ifnum #1\expandafter \@firstoftwo
 \else \expandafter \@secondoftwo
 \fi
}%
\providecommand \@ifx [1]{%
 \ifx #1\expandafter \@firstoftwo
 \else \expandafter \@secondoftwo
 \fi
}%
\providecommand \natexlab [1]{#1}%
\providecommand \enquote  [1]{``#1''}%
\providecommand \bibnamefont  [1]{#1}%
\providecommand \bibfnamefont [1]{#1}%
\providecommand \citenamefont [1]{#1}%
\providecommand \href@noop [0]{\@secondoftwo}%
\providecommand \href [0]{\begingroup \@sanitize@url \@href}%
\providecommand \@href[1]{\@@startlink{#1}\@@href}%
\providecommand \@@href[1]{\endgroup#1\@@endlink}%
\providecommand \@sanitize@url [0]{\catcode `\\12\catcode `\$12\catcode
  `\&12\catcode `\#12\catcode `\^12\catcode `\_12\catcode `\%12\relax}%
\providecommand \@@startlink[1]{}%
\providecommand \@@endlink[0]{}%
\providecommand \url  [0]{\begingroup\@sanitize@url \@url }%
\providecommand \@url [1]{\endgroup\@href {#1}{\urlprefix }}%
\providecommand \urlprefix  [0]{URL }%
\providecommand \Eprint [0]{\href }%
\providecommand \doibase [0]{http://dx.doi.org/}%
\providecommand \selectlanguage [0]{\@gobble}%
\providecommand \bibinfo  [0]{\@secondoftwo}%
\providecommand \bibfield  [0]{\@secondoftwo}%
\providecommand \translation [1]{[#1]}%
\providecommand \BibitemOpen [0]{}%
\providecommand \bibitemStop [0]{}%
\providecommand \bibitemNoStop [0]{.\EOS\space}%
\providecommand \EOS [0]{\spacefactor3000\relax}%
\providecommand \BibitemShut  [1]{\csname bibitem#1\endcsname}%
\let\auto@bib@innerbib\@empty
\bibitem [{\citenamefont {Tucker}\ \emph {et~al.}(2005)\citenamefont {Tucker},
  \citenamefont {Goodwin}, \citenamefont {Dove}, \citenamefont {Keen},
  \citenamefont {Wells},\ and\ \citenamefont {Evans}}]{tucker2005negative}%
  \BibitemOpen
  \bibfield  {author} {\bibinfo {author} {\bibfnamefont {M.~G.}\ \bibnamefont
  {Tucker}}, \bibinfo {author} {\bibfnamefont {A.~L.}\ \bibnamefont {Goodwin}},
  \bibinfo {author} {\bibfnamefont {M.~T.}\ \bibnamefont {Dove}}, \bibinfo
  {author} {\bibfnamefont {D.~A.}\ \bibnamefont {Keen}}, \bibinfo {author}
  {\bibfnamefont {S.~A.}\ \bibnamefont {Wells}}, \ and\ \bibinfo {author}
  {\bibfnamefont {J.~S.}\ \bibnamefont {Evans}},\ }\href@noop {} {\bibfield
  {journal} {\bibinfo  {journal} {Phys. Rev. Lett.}\ }\textbf {\bibinfo
  {volume} {95}},\ \bibinfo {pages} {255501} (\bibinfo {year}
  {2005})}\BibitemShut {NoStop}%
\bibitem [{\citenamefont {Mounet}\ and\ \citenamefont
  {Marzari}(2005)}]{mounet05}%
  \BibitemOpen
  \bibfield  {author} {\bibinfo {author} {\bibfnamefont {N.}~\bibnamefont
  {Mounet}}\ and\ \bibinfo {author} {\bibfnamefont {N.}~\bibnamefont
  {Marzari}},\ }\href@noop {} {\bibfield  {journal} {\bibinfo  {journal} {Phys.
  Rev. B}\ }\textbf {\bibinfo {volume} {71}},\ \bibinfo {pages} {205214}
  (\bibinfo {year} {2005})}\BibitemShut {NoStop}%
\bibitem [{\citenamefont {Kim}\ \emph {et~al.}(2018)\citenamefont {Kim},
  \citenamefont {Hellman}, \citenamefont {Herriman}, \citenamefont {Smith},
  \citenamefont {Lin}, \citenamefont {Shulumba}, \citenamefont {Niedziela},
  \citenamefont {Li}, \citenamefont {Abernathy},\ and\ \citenamefont
  {Fultz}}]{kim2018nuclear}%
  \BibitemOpen
  \bibfield  {author} {\bibinfo {author} {\bibfnamefont {D.~S.}\ \bibnamefont
  {Kim}}, \bibinfo {author} {\bibfnamefont {O.}~\bibnamefont {Hellman}},
  \bibinfo {author} {\bibfnamefont {J.}~\bibnamefont {Herriman}}, \bibinfo
  {author} {\bibfnamefont {H.}~\bibnamefont {Smith}}, \bibinfo {author}
  {\bibfnamefont {J.}~\bibnamefont {Lin}}, \bibinfo {author} {\bibfnamefont
  {N.}~\bibnamefont {Shulumba}}, \bibinfo {author} {\bibfnamefont
  {J.}~\bibnamefont {Niedziela}}, \bibinfo {author} {\bibfnamefont
  {C.}~\bibnamefont {Li}}, \bibinfo {author} {\bibfnamefont {D.}~\bibnamefont
  {Abernathy}}, \ and\ \bibinfo {author} {\bibfnamefont {B.}~\bibnamefont
  {Fultz}},\ }\href@noop {} {\bibfield  {journal} {\bibinfo  {journal} {Proc.
  Natl. Acad. Sci. U.S.A.}\ }\textbf {\bibinfo {volume} {115}},\ \bibinfo
  {pages} {1992} (\bibinfo {year} {2018})}\BibitemShut {NoStop}%
\bibitem [{\citenamefont {Ritz}\ and\ \citenamefont
  {Benedek}(2018)}]{ritz2018interplay}%
  \BibitemOpen
  \bibfield  {author} {\bibinfo {author} {\bibfnamefont {E.~T.}\ \bibnamefont
  {Ritz}}\ and\ \bibinfo {author} {\bibfnamefont {N.~A.}\ \bibnamefont
  {Benedek}},\ }\href@noop {} {\bibfield  {journal} {\bibinfo  {journal} {Phys.
  Rev. Lett.}\ }\textbf {\bibinfo {volume} {121}},\ \bibinfo {pages} {255901}
  (\bibinfo {year} {2018})}\BibitemShut {NoStop}%
\bibitem [{\citenamefont {Goodwin}\ \emph
  {et~al.}(2008{\natexlab{a}})\citenamefont {Goodwin}, \citenamefont {Keen},\
  and\ \citenamefont {Tucker}}]{goodwin08}%
  \BibitemOpen
  \bibfield  {author} {\bibinfo {author} {\bibfnamefont {A.~L.}\ \bibnamefont
  {Goodwin}}, \bibinfo {author} {\bibfnamefont {D.~A.}\ \bibnamefont {Keen}}, \
  and\ \bibinfo {author} {\bibfnamefont {M.~G.}\ \bibnamefont {Tucker}},\
  }\href@noop {} {\bibfield  {journal} {\bibinfo  {journal} {Proc. Natl. Acad.
  Sci. U.S.A.}\ }\textbf {\bibinfo {volume} {105}},\ \bibinfo {pages} {18708}
  (\bibinfo {year} {2008}{\natexlab{a}})}\BibitemShut {NoStop}%
\bibitem [{\citenamefont {Goodwin}\ \emph
  {et~al.}(2008{\natexlab{b}})\citenamefont {Goodwin}, \citenamefont {Keen},
  \citenamefont {Tucker}, \citenamefont {Dove}, \citenamefont {Peters},\ and\
  \citenamefont {Evans}}]{goodwin08b}%
  \BibitemOpen
  \bibfield  {author} {\bibinfo {author} {\bibfnamefont {A.~L.}\ \bibnamefont
  {Goodwin}}, \bibinfo {author} {\bibfnamefont {D.~A.}\ \bibnamefont {Keen}},
  \bibinfo {author} {\bibfnamefont {M.~G.}\ \bibnamefont {Tucker}}, \bibinfo
  {author} {\bibfnamefont {M.~T.}\ \bibnamefont {Dove}}, \bibinfo {author}
  {\bibfnamefont {L.}~\bibnamefont {Peters}}, \ and\ \bibinfo {author}
  {\bibfnamefont {J.~S.~O.}\ \bibnamefont {Evans}},\ }\href@noop {} {\bibfield
  {journal} {\bibinfo  {journal} {J. Am. Chem. Soc.}\ }\textbf {\bibinfo
  {volume} {130}},\ \bibinfo {pages} {9660} (\bibinfo {year}
  {2008}{\natexlab{b}})}\BibitemShut {NoStop}%
\bibitem [{\citenamefont {Goodwin}\ \emph {et~al.}(2009)\citenamefont
  {Goodwin}, \citenamefont {Kennedy},\ and\ \citenamefont
  {Kepert}}]{goodwin09}%
  \BibitemOpen
  \bibfield  {author} {\bibinfo {author} {\bibfnamefont {A.~L.}\ \bibnamefont
  {Goodwin}}, \bibinfo {author} {\bibfnamefont {B.~J.}\ \bibnamefont
  {Kennedy}}, \ and\ \bibinfo {author} {\bibfnamefont {C.~J.}\ \bibnamefont
  {Kepert}},\ }\href@noop {} {\bibfield  {journal} {\bibinfo  {journal} {J. Am.
  Chem. Soc.}\ }\textbf {\bibinfo {volume} {131}},\ \bibinfo {pages} {6334}
  (\bibinfo {year} {2009})}\BibitemShut {NoStop}%
\bibitem [{\citenamefont {Lind}(2012)}]{lind2012two}%
  \BibitemOpen
  \bibfield  {author} {\bibinfo {author} {\bibfnamefont {C.}~\bibnamefont
  {Lind}},\ }\href {\doibase 10.3390/ma5061125} {\bibfield  {journal} {\bibinfo
   {journal} {Materials}\ }\textbf {\bibinfo {volume} {5}},\ \bibinfo {pages}
  {1125} (\bibinfo {year} {2012})}\BibitemShut {NoStop}%
\bibitem [{\citenamefont {Miller}\ \emph {et~al.}(2009)\citenamefont {Miller},
  \citenamefont {Smith}, \citenamefont {Mackenzie},\ and\ \citenamefont
  {Evans}}]{miller2009negative}%
  \BibitemOpen
  \bibfield  {author} {\bibinfo {author} {\bibfnamefont {W.}~\bibnamefont
  {Miller}}, \bibinfo {author} {\bibfnamefont {C.}~\bibnamefont {Smith}},
  \bibinfo {author} {\bibfnamefont {D.}~\bibnamefont {Mackenzie}}, \ and\
  \bibinfo {author} {\bibfnamefont {K.}~\bibnamefont {Evans}},\ }\href@noop {}
  {\bibfield  {journal} {\bibinfo  {journal} {J. Mater. Sci.}\ }\textbf
  {\bibinfo {volume} {44}},\ \bibinfo {pages} {5441} (\bibinfo {year}
  {2009})}\BibitemShut {NoStop}%
\bibitem [{\citenamefont {Barrera}\ \emph {et~al.}(2005)\citenamefont
  {Barrera}, \citenamefont {Bruno}, \citenamefont {Barron},\ and\ \citenamefont
  {Allan}}]{barrera2005negative}%
  \BibitemOpen
  \bibfield  {author} {\bibinfo {author} {\bibfnamefont {G.~D.}\ \bibnamefont
  {Barrera}}, \bibinfo {author} {\bibfnamefont {J.~A.~O.}\ \bibnamefont
  {Bruno}}, \bibinfo {author} {\bibfnamefont {T.~H.~K.}\ \bibnamefont
  {Barron}}, \ and\ \bibinfo {author} {\bibfnamefont {N.~L.}\ \bibnamefont
  {Allan}},\ }\href@noop {} {\bibfield  {journal} {\bibinfo  {journal} {J.
  Phys. Condens. Matter}\ }\textbf {\bibinfo {volume} {17}},\ \bibinfo {pages}
  {R217} (\bibinfo {year} {2005})}\BibitemShut {NoStop}%
\bibitem [{\citenamefont {Yanase}\ \emph {et~al.}(2011)\citenamefont {Yanase},
  \citenamefont {Kojima},\ and\ \citenamefont {Kobayashi}}]{yanase2011effects}%
  \BibitemOpen
  \bibfield  {author} {\bibinfo {author} {\bibfnamefont {I.}~\bibnamefont
  {Yanase}}, \bibinfo {author} {\bibfnamefont {T.}~\bibnamefont {Kojima}}, \
  and\ \bibinfo {author} {\bibfnamefont {H.}~\bibnamefont {Kobayashi}},\
  }\href@noop {} {\bibfield  {journal} {\bibinfo  {journal} {Solid State
  Commun.}\ }\textbf {\bibinfo {volume} {151}},\ \bibinfo {pages} {595}
  (\bibinfo {year} {2011})}\BibitemShut {NoStop}%
\bibitem [{\citenamefont {Ablitt}\ \emph {et~al.}(2019)\citenamefont {Ablitt},
  \citenamefont {McCay}, \citenamefont {Craddock}, \citenamefont {Cooper},
  \citenamefont {Reynolds}, \citenamefont {Mostofi}, \citenamefont {Bristowe},
  \citenamefont {Murray},\ and\ \citenamefont {Senn}}]{ablitt2019tolerance}%
  \BibitemOpen
  \bibfield  {author} {\bibinfo {author} {\bibfnamefont {C.}~\bibnamefont
  {Ablitt}}, \bibinfo {author} {\bibfnamefont {H.}~\bibnamefont {McCay}},
  \bibinfo {author} {\bibfnamefont {S.}~\bibnamefont {Craddock}}, \bibinfo
  {author} {\bibfnamefont {L.}~\bibnamefont {Cooper}}, \bibinfo {author}
  {\bibfnamefont {E.}~\bibnamefont {Reynolds}}, \bibinfo {author}
  {\bibfnamefont {A.~A.}\ \bibnamefont {Mostofi}}, \bibinfo {author}
  {\bibfnamefont {N.~C.}\ \bibnamefont {Bristowe}}, \bibinfo {author}
  {\bibfnamefont {C.~A.}\ \bibnamefont {Murray}}, \ and\ \bibinfo {author}
  {\bibfnamefont {M.~S.}\ \bibnamefont {Senn}},\ }\href@noop {} {\bibfield
  {journal} {\bibinfo  {journal} {Chem. Mater.}\ } (\bibinfo {year}
  {2019})}\BibitemShut {NoStop}%
\bibitem [{\citenamefont {Chen}\ \emph {et~al.}(2005)\citenamefont {Chen},
  \citenamefont {Xing}, \citenamefont {Yu},\ and\ \citenamefont
  {Liu}}]{chen2005thermal}%
  \BibitemOpen
  \bibfield  {author} {\bibinfo {author} {\bibfnamefont {J.}~\bibnamefont
  {Chen}}, \bibinfo {author} {\bibfnamefont {X.}~\bibnamefont {Xing}}, \bibinfo
  {author} {\bibfnamefont {R.}~\bibnamefont {Yu}}, \ and\ \bibinfo {author}
  {\bibfnamefont {G.}~\bibnamefont {Liu}},\ }\href@noop {} {\bibfield
  {journal} {\bibinfo  {journal} {J. Am. Ceram. Soc.}\ }\textbf {\bibinfo
  {volume} {88}},\ \bibinfo {pages} {1356} (\bibinfo {year}
  {2005})}\BibitemShut {NoStop}%
\bibitem [{\citenamefont {Shirane}\ and\ \citenamefont
  {Hoshino}(1951)}]{shirane1951phase}%
  \BibitemOpen
  \bibfield  {author} {\bibinfo {author} {\bibfnamefont {G.}~\bibnamefont
  {Shirane}}\ and\ \bibinfo {author} {\bibfnamefont {S.}~\bibnamefont
  {Hoshino}},\ }\href@noop {} {\bibfield  {journal} {\bibinfo  {journal} {J.
  Phys. Soc. Jpn.}\ }\textbf {\bibinfo {volume} {6}},\ \bibinfo {pages} {265}
  (\bibinfo {year} {1951})}\BibitemShut {NoStop}%
\bibitem [{\citenamefont {Janolin}\ \emph {et~al.}(2007)\citenamefont
  {Janolin}, \citenamefont {Le~Marrec}, \citenamefont {Chevreul},\ and\
  \citenamefont {Dkhil}}]{janolin2007temperature}%
  \BibitemOpen
  \bibfield  {author} {\bibinfo {author} {\bibfnamefont {P.-E.}\ \bibnamefont
  {Janolin}}, \bibinfo {author} {\bibfnamefont {F.}~\bibnamefont {Le~Marrec}},
  \bibinfo {author} {\bibfnamefont {J.}~\bibnamefont {Chevreul}}, \ and\
  \bibinfo {author} {\bibfnamefont {B.}~\bibnamefont {Dkhil}},\ }\href@noop {}
  {\bibfield  {journal} {\bibinfo  {journal} {Appl. Phys. Lett.}\ }\textbf
  {\bibinfo {volume} {90}},\ \bibinfo {pages} {192910} (\bibinfo {year}
  {2007})}\BibitemShut {NoStop}%
\bibitem [{\citenamefont {Janolin}(2009)}]{janolin09}%
  \BibitemOpen
  \bibfield  {author} {\bibinfo {author} {\bibfnamefont {P.-E.}\ \bibnamefont
  {Janolin}},\ }\href {\doibase 10.1007/s10853-009-3553-1} {\bibfield
  {journal} {\bibinfo  {journal} {Journal of Materials Science}\ }\textbf
  {\bibinfo {volume} {44}},\ \bibinfo {pages} {5025} (\bibinfo {year}
  {2009})}\BibitemShut {NoStop}%
\bibitem [{\citenamefont {Giannozzi}\ \emph {et~al.}(2009)\citenamefont
  {Giannozzi}, \citenamefont {Baroni}, \citenamefont {Bonini}, \citenamefont
  {Calandra}, \citenamefont {Car}, \citenamefont {Cavazzoni}, \citenamefont
  {Ceresoli}, \citenamefont {Chiarotti}, \citenamefont {Cococcioni},
  \citenamefont {Dabo} \emph {et~al.}}]{giannozzi2009quantum}%
  \BibitemOpen
  \bibfield  {author} {\bibinfo {author} {\bibfnamefont {P.}~\bibnamefont
  {Giannozzi}}, \bibinfo {author} {\bibfnamefont {S.}~\bibnamefont {Baroni}},
  \bibinfo {author} {\bibfnamefont {N.}~\bibnamefont {Bonini}}, \bibinfo
  {author} {\bibfnamefont {M.}~\bibnamefont {Calandra}}, \bibinfo {author}
  {\bibfnamefont {R.}~\bibnamefont {Car}}, \bibinfo {author} {\bibfnamefont
  {C.}~\bibnamefont {Cavazzoni}}, \bibinfo {author} {\bibfnamefont
  {D.}~\bibnamefont {Ceresoli}}, \bibinfo {author} {\bibfnamefont {G.~L.}\
  \bibnamefont {Chiarotti}}, \bibinfo {author} {\bibfnamefont {M.}~\bibnamefont
  {Cococcioni}}, \bibinfo {author} {\bibfnamefont {I.}~\bibnamefont {Dabo}},
  \emph {et~al.},\ }\href@noop {} {\bibfield  {journal} {\bibinfo  {journal}
  {J. Phys. Condens. Matter}\ }\textbf {\bibinfo {volume} {21}},\ \bibinfo
  {pages} {395502} (\bibinfo {year} {2009})}\BibitemShut {NoStop}%
\bibitem [{\citenamefont {Wu}\ and\ \citenamefont {Cohen}(2006)}]{WC2006}%
  \BibitemOpen
  \bibfield  {author} {\bibinfo {author} {\bibfnamefont {Z.}~\bibnamefont
  {Wu}}\ and\ \bibinfo {author} {\bibfnamefont {R.~E.}\ \bibnamefont {Cohen}},\
  }\href@noop {} {\bibfield  {journal} {\bibinfo  {journal} {Phys. Rev. B}\
  }\textbf {\bibinfo {volume} {73}},\ \bibinfo {pages} {235116} (\bibinfo
  {year} {2006})}\BibitemShut {NoStop}%
\bibitem [{\citenamefont {Garrity}\ \emph {et~al.}(2014)\citenamefont
  {Garrity}, \citenamefont {Bennett}, \citenamefont {Rabe},\ and\ \citenamefont
  {Vanderbilt}}]{garrity2014pseudopotentials}%
  \BibitemOpen
  \bibfield  {author} {\bibinfo {author} {\bibfnamefont {K.~F.}\ \bibnamefont
  {Garrity}}, \bibinfo {author} {\bibfnamefont {J.~W.}\ \bibnamefont
  {Bennett}}, \bibinfo {author} {\bibfnamefont {K.~M.}\ \bibnamefont {Rabe}}, \
  and\ \bibinfo {author} {\bibfnamefont {D.}~\bibnamefont {Vanderbilt}},\
  }\href@noop {} {\bibfield  {journal} {\bibinfo  {journal} {Comput. Mater.
  Sci.}\ }\textbf {\bibinfo {volume} {81}},\ \bibinfo {pages} {446} (\bibinfo
  {year} {2014})}\BibitemShut {NoStop}%
\bibitem [{\citenamefont {Ritz}\ \emph {et~al.}(2019)\citenamefont {Ritz},
  \citenamefont {Li},\ and\ \citenamefont {Benedek}}]{ritz2019thermal}%
  \BibitemOpen
  \bibfield  {author} {\bibinfo {author} {\bibfnamefont {E.~T.}\ \bibnamefont
  {Ritz}}, \bibinfo {author} {\bibfnamefont {S.~J.}\ \bibnamefont {Li}}, \ and\
  \bibinfo {author} {\bibfnamefont {N.~A.}\ \bibnamefont {Benedek}},\
  }\href@noop {} {\bibfield  {journal} {\bibinfo  {journal} {J. Appl. Phys.}\
  }\textbf {\bibinfo {volume} {126}},\ \bibinfo {pages} {171102} (\bibinfo
  {year} {2019})}\BibitemShut {NoStop}%
\bibitem [{\citenamefont {Ritz}\ and\ \citenamefont
  {Benedek}(2020)}]{ritz2020strain}%
  \BibitemOpen
  \bibfield  {author} {\bibinfo {author} {\bibfnamefont {E.~T.}\ \bibnamefont
  {Ritz}}\ and\ \bibinfo {author} {\bibfnamefont {N.~A.}\ \bibnamefont
  {Benedek}},\ }\href@noop {} {\bibfield  {journal} {\bibinfo  {journal}
  {Physical Review Materials}\ }\textbf {\bibinfo {volume} {4}},\ \bibinfo
  {pages} {084410} (\bibinfo {year} {2020})}\BibitemShut {NoStop}%
\bibitem [{ins()}]{instab}%
  \BibitemOpen
  \href@noop {} {}\bibinfo {note} {Our DFT simulations predict that the unit
  cell of bulk PbTiO$_3$ exhibits unstable phonon modes at 0 K in its cubic
  phase, as well as for nearly-cubic unit cells with very low $c/a$ ratios.
  This results in difficulty when calculating QHA grid points very close to the
  tetragonal-cubic phase transition. The lowest value of $c$ in the set of unit
  cells used in our quasiharmonic grid is approximately 4.00 \AA, meaning that
  for temperatures greater than 720 K, the point that minimizes the Helmholtz
  free energy surface fit is no longer bounded by that grid along the $c$ axis.
  While such an extrapolation of the derivatives of the vibrational free energy
  is within the spirit of the Gru\"neisen theory of thermal expansion, our
  simulation can be interpreted to conclude that the tetragonal-cubic phase
  transition occurs somewhere within the range of 720 K to 800 K.}\BibitemShut
  {Stop}%
\bibitem [{\citenamefont {Wallace}(1972)}]{wallace}%
  \BibitemOpen
  \bibfield  {author} {\bibinfo {author} {\bibfnamefont {D.~C.}\ \bibnamefont
  {Wallace}},\ }\href@noop {} {\emph {\bibinfo {title} {Thermodynamics of
  Crystals}}}\ (\bibinfo  {publisher} {Wiley},\ \bibinfo {year}
  {1972})\BibitemShut {NoStop}%
\bibitem [{\citenamefont {Horton}\ and\ \citenamefont
  {Maradudin}(1974)}]{horton1974dynamical}%
  \BibitemOpen
  \bibfield  {author} {\bibinfo {author} {\bibfnamefont {G.~K.}\ \bibnamefont
  {Horton}}\ and\ \bibinfo {author} {\bibfnamefont {A.~A.}\ \bibnamefont
  {Maradudin}},\ }\href@noop {} {\emph {\bibinfo {title} {{Dynamical Properties
  of Solids}}}},\ Vol.~\bibinfo {volume} {1}\ (\bibinfo  {publisher}
  {Elsevier},\ \bibinfo {year} {1974})\BibitemShut {NoStop}%
\bibitem [{\citenamefont {Choy}\ \emph {et~al.}(1984)\citenamefont {Choy},
  \citenamefont {Wong},\ and\ \citenamefont {Young}}]{choy1984thermal}%
  \BibitemOpen
  \bibfield  {author} {\bibinfo {author} {\bibfnamefont {C.}~\bibnamefont
  {Choy}}, \bibinfo {author} {\bibfnamefont {S.}~\bibnamefont {Wong}}, \ and\
  \bibinfo {author} {\bibfnamefont {K.}~\bibnamefont {Young}},\ }\href@noop {}
  {\bibfield  {journal} {\bibinfo  {journal} {Phys. Rev. B}\ }\textbf {\bibinfo
  {volume} {29}},\ \bibinfo {pages} {1741} (\bibinfo {year}
  {1984})}\BibitemShut {NoStop}%
\bibitem [{\citenamefont {Ashcroft}\ and\ \citenamefont
  {Mermin}(1976)}]{ashcroft2005solid}%
  \BibitemOpen
  \bibfield  {author} {\bibinfo {author} {\bibfnamefont {N.~W.}\ \bibnamefont
  {Ashcroft}}\ and\ \bibinfo {author} {\bibfnamefont {N.~D.}\ \bibnamefont
  {Mermin}},\ }\href@noop {} {\emph {\bibinfo {title} {Solid State Physics}}}\
  (\bibinfo  {publisher} {Holt, Rinehart and Winston, New York},\ \bibinfo
  {year} {1976})\BibitemShut {NoStop}%
\bibitem [{\citenamefont {Munn}(1972)}]{munn1972role}%
  \BibitemOpen
  \bibfield  {author} {\bibinfo {author} {\bibfnamefont {R.}~\bibnamefont
  {Munn}},\ }\href@noop {} {\bibfield  {journal} {\bibinfo  {journal} {J. Phys.
  C: Solid State Physics}\ }\textbf {\bibinfo {volume} {5}},\ \bibinfo {pages}
  {535} (\bibinfo {year} {1972})}\BibitemShut {NoStop}%
\bibitem [{\citenamefont {Barron}\ and\ \citenamefont
  {Klein}(1965)}]{barron1965second}%
  \BibitemOpen
  \bibfield  {author} {\bibinfo {author} {\bibfnamefont {T.}~\bibnamefont
  {Barron}}\ and\ \bibinfo {author} {\bibfnamefont {M.}~\bibnamefont {Klein}},\
  }\href@noop {} {\bibfield  {journal} {\bibinfo  {journal} {Proceedings of the
  Physical Society (1958-1967)}\ }\textbf {\bibinfo {volume} {85}},\ \bibinfo
  {pages} {523} (\bibinfo {year} {1965})}\BibitemShut {NoStop}%
\bibitem [{\citenamefont {Wallace}(1965)}]{wallace1965lattice}%
  \BibitemOpen
  \bibfield  {author} {\bibinfo {author} {\bibfnamefont {D.~C.}\ \bibnamefont
  {Wallace}},\ }\href@noop {} {\bibfield  {journal} {\bibinfo  {journal}
  {Reviews of Modern Physics}\ }\textbf {\bibinfo {volume} {37}},\ \bibinfo
  {pages} {57} (\bibinfo {year} {1965})}\BibitemShut {NoStop}%
\bibitem [{\citenamefont {Wang}\ and\ \citenamefont
  {Li}(2012)}]{wang2012nonlinear}%
  \BibitemOpen
  \bibfield  {author} {\bibinfo {author} {\bibfnamefont {H.}~\bibnamefont
  {Wang}}\ and\ \bibinfo {author} {\bibfnamefont {M.}~\bibnamefont {Li}},\
  }\href@noop {} {\bibfield  {journal} {\bibinfo  {journal} {Physical Review
  B}\ }\textbf {\bibinfo {volume} {85}},\ \bibinfo {pages} {104103} (\bibinfo
  {year} {2012})}\BibitemShut {NoStop}%
\bibitem [{str()}]{strain}%
  \BibitemOpen
  \href@noop {} {}\bibinfo {note} {The films under compressive and tensile
  strain at 300~K remain under compressive and tensile strain throughout nearly
  the entire temperature range studied (the only exception is the film
  constrained by the substrate with $\varepsilon_a^{300}=+0.75\%$ at $\alpha_a$
  = 1$\times$10$^{-5}$ for temperatures over 700~K).}\BibitemShut {Stop}%
\bibitem [{\citenamefont {Chakoumakos}\ \emph {et~al.}(1998)\citenamefont
  {Chakoumakos}, \citenamefont {Schlom}, \citenamefont {Urbanik},\ and\
  \citenamefont {Luine}}]{chakoumakos1998thermal}%
  \BibitemOpen
  \bibfield  {author} {\bibinfo {author} {\bibfnamefont {B.}~\bibnamefont
  {Chakoumakos}}, \bibinfo {author} {\bibfnamefont {D.}~\bibnamefont {Schlom}},
  \bibinfo {author} {\bibfnamefont {M.}~\bibnamefont {Urbanik}}, \ and\
  \bibinfo {author} {\bibfnamefont {J.}~\bibnamefont {Luine}},\ }\href@noop {}
  {\bibfield  {journal} {\bibinfo  {journal} {J. Appl. Phys.}\ }\textbf
  {\bibinfo {volume} {83}},\ \bibinfo {pages} {1979} (\bibinfo {year}
  {1998})}\BibitemShut {NoStop}%
\bibitem [{\citenamefont {de~Ligny}\ and\ \citenamefont
  {Richet}(1996)}]{de1996high}%
  \BibitemOpen
  \bibfield  {author} {\bibinfo {author} {\bibfnamefont {D.}~\bibnamefont
  {de~Ligny}}\ and\ \bibinfo {author} {\bibfnamefont {P.}~\bibnamefont
  {Richet}},\ }\href@noop {} {\bibfield  {journal} {\bibinfo  {journal} {Phys.
  Rev. B}\ }\textbf {\bibinfo {volume} {53}},\ \bibinfo {pages} {3013}
  (\bibinfo {year} {1996})}\BibitemShut {NoStop}%
\bibitem [{\citenamefont {Biegalski}\ \emph {et~al.}(2005)\citenamefont
  {Biegalski}, \citenamefont {Haeni}, \citenamefont {Trolier-McKinstry},
  \citenamefont {Schlom}, \citenamefont {Brandle},\ and\ \citenamefont
  {Graitis}}]{biegalski2005thermal}%
  \BibitemOpen
  \bibfield  {author} {\bibinfo {author} {\bibfnamefont {M.}~\bibnamefont
  {Biegalski}}, \bibinfo {author} {\bibfnamefont {J.}~\bibnamefont {Haeni}},
  \bibinfo {author} {\bibfnamefont {S.}~\bibnamefont {Trolier-McKinstry}},
  \bibinfo {author} {\bibfnamefont {D.}~\bibnamefont {Schlom}}, \bibinfo
  {author} {\bibfnamefont {C.}~\bibnamefont {Brandle}}, \ and\ \bibinfo
  {author} {\bibfnamefont {A.~V.}\ \bibnamefont {Graitis}},\ }\href@noop {}
  {\bibfield  {journal} {\bibinfo  {journal} {J. Mater. Res.}\ }\textbf
  {\bibinfo {volume} {20}},\ \bibinfo {pages} {952} (\bibinfo {year}
  {2005})}\BibitemShut {NoStop}%
\bibitem [{\citenamefont {Dangi{\'c}}\ \emph {et~al.}(2018)\citenamefont
  {Dangi{\'c}}, \citenamefont {Murphy}, \citenamefont {Murray}, \citenamefont
  {Fahy},\ and\ \citenamefont {Savi{\'c}}}]{dangic2018coupling}%
  \BibitemOpen
  \bibfield  {author} {\bibinfo {author} {\bibfnamefont {D.}~\bibnamefont
  {Dangi{\'c}}}, \bibinfo {author} {\bibfnamefont {A.~R.}\ \bibnamefont
  {Murphy}}, \bibinfo {author} {\bibfnamefont {{\'E}.~D.}\ \bibnamefont
  {Murray}}, \bibinfo {author} {\bibfnamefont {S.}~\bibnamefont {Fahy}}, \ and\
  \bibinfo {author} {\bibfnamefont {I.}~\bibnamefont {Savi{\'c}}},\ }\href@noop
  {} {\bibfield  {journal} {\bibinfo  {journal} {Physical Review B}\ }\textbf
  {\bibinfo {volume} {97}},\ \bibinfo {pages} {224106} (\bibinfo {year}
  {2018})}\BibitemShut {NoStop}%
\bibitem [{\citenamefont {Tachibana}\ \emph {et~al.}(2008)\citenamefont
  {Tachibana}, \citenamefont {Kolodiazhnyi},\ and\ \citenamefont
  {Takayama-Muromachi}}]{tachibana2008thermal}%
  \BibitemOpen
  \bibfield  {author} {\bibinfo {author} {\bibfnamefont {M.}~\bibnamefont
  {Tachibana}}, \bibinfo {author} {\bibfnamefont {T.}~\bibnamefont
  {Kolodiazhnyi}}, \ and\ \bibinfo {author} {\bibfnamefont {E.}~\bibnamefont
  {Takayama-Muromachi}},\ }\href@noop {} {\bibfield  {journal} {\bibinfo
  {journal} {Applied Physics Letters}\ }\textbf {\bibinfo {volume} {93}},\
  \bibinfo {pages} {092902} (\bibinfo {year} {2008})}\BibitemShut {NoStop}%
\bibitem [{\citenamefont {Yoshida}(1960)}]{yoshida1960thermal}%
  \BibitemOpen
  \bibfield  {author} {\bibinfo {author} {\bibfnamefont {I.}~\bibnamefont
  {Yoshida}},\ }\href@noop {} {\bibfield  {journal} {\bibinfo  {journal}
  {Journal of the Physical Society of Japan}\ }\textbf {\bibinfo {volume}
  {15}},\ \bibinfo {pages} {2211} (\bibinfo {year} {1960})}\BibitemShut
  {NoStop}%
\bibitem [{\citenamefont {Pertsev}\ \emph {et~al.}(1998)\citenamefont
  {Pertsev}, \citenamefont {Zembilgotov},\ and\ \citenamefont
  {Tagantsev}}]{pertsev1998effect}%
  \BibitemOpen
  \bibfield  {author} {\bibinfo {author} {\bibfnamefont {N.}~\bibnamefont
  {Pertsev}}, \bibinfo {author} {\bibfnamefont {A.}~\bibnamefont
  {Zembilgotov}}, \ and\ \bibinfo {author} {\bibfnamefont {A.}~\bibnamefont
  {Tagantsev}},\ }\href@noop {} {\bibfield  {journal} {\bibinfo  {journal}
  {Phys. Rev. Lett.}\ }\textbf {\bibinfo {volume} {80}},\ \bibinfo {pages}
  {1988} (\bibinfo {year} {1998})}\BibitemShut {NoStop}%
\bibitem [{\citenamefont {Zembilgotov}\ \emph {et~al.}(2002)\citenamefont
  {Zembilgotov}, \citenamefont {Pertsev}, \citenamefont {Kohlstedt},\ and\
  \citenamefont {Waser}}]{waser02}%
  \BibitemOpen
  \bibfield  {author} {\bibinfo {author} {\bibfnamefont {A.~G.}\ \bibnamefont
  {Zembilgotov}}, \bibinfo {author} {\bibfnamefont {N.~A.}\ \bibnamefont
  {Pertsev}}, \bibinfo {author} {\bibfnamefont {H.}~\bibnamefont {Kohlstedt}},
  \ and\ \bibinfo {author} {\bibfnamefont {R.}~\bibnamefont {Waser}},\ }\href
  {\doibase 10.1063/1.1427406} {\bibfield  {journal} {\bibinfo  {journal}
  {Journal of Applied Physics}\ }\textbf {\bibinfo {volume} {91}},\ \bibinfo
  {pages} {2247} (\bibinfo {year} {2002})},\ \Eprint
  {http://arxiv.org/abs/cond-mat/0111218} {cond-mat/0111218} \BibitemShut
  {NoStop}%
\bibitem [{\citenamefont {Highland}\ \emph {et~al.}(2014)\citenamefont
  {Highland}, \citenamefont {Fong}, \citenamefont {Stephenson}, \citenamefont
  {Fister}, \citenamefont {Fuoss}, \citenamefont {Streiffer}, \citenamefont
  {Thompson}, \citenamefont {Richard},\ and\ \citenamefont
  {Eastman}}]{highland2014interfacial}%
  \BibitemOpen
  \bibfield  {author} {\bibinfo {author} {\bibfnamefont {M.}~\bibnamefont
  {Highland}}, \bibinfo {author} {\bibfnamefont {D.}~\bibnamefont {Fong}},
  \bibinfo {author} {\bibfnamefont {G.}~\bibnamefont {Stephenson}}, \bibinfo
  {author} {\bibfnamefont {T.}~\bibnamefont {Fister}}, \bibinfo {author}
  {\bibfnamefont {P.}~\bibnamefont {Fuoss}}, \bibinfo {author} {\bibfnamefont
  {S.}~\bibnamefont {Streiffer}}, \bibinfo {author} {\bibfnamefont
  {C.}~\bibnamefont {Thompson}}, \bibinfo {author} {\bibfnamefont {M.-I.}\
  \bibnamefont {Richard}}, \ and\ \bibinfo {author} {\bibfnamefont
  {J.}~\bibnamefont {Eastman}},\ }\href@noop {} {\bibfield  {journal} {\bibinfo
   {journal} {Appl. Phys. Lett.}\ }\textbf {\bibinfo {volume} {104}},\ \bibinfo
  {pages} {132901} (\bibinfo {year} {2014})}\BibitemShut {NoStop}%
\bibitem [{\citenamefont {Lichtensteiger}\ \emph {et~al.}(2005)\citenamefont
  {Lichtensteiger}, \citenamefont {Triscone}, \citenamefont {Junquera},\ and\
  \citenamefont {Ghosez}}]{lichtensteiger2005ferroelectricity}%
  \BibitemOpen
  \bibfield  {author} {\bibinfo {author} {\bibfnamefont {C.}~\bibnamefont
  {Lichtensteiger}}, \bibinfo {author} {\bibfnamefont {J.-M.}\ \bibnamefont
  {Triscone}}, \bibinfo {author} {\bibfnamefont {J.}~\bibnamefont {Junquera}},
  \ and\ \bibinfo {author} {\bibfnamefont {P.}~\bibnamefont {Ghosez}},\
  }\href@noop {} {\bibfield  {journal} {\bibinfo  {journal} {Physical review
  letters}\ }\textbf {\bibinfo {volume} {94}},\ \bibinfo {pages} {047603}
  (\bibinfo {year} {2005})}\BibitemShut {NoStop}%
\bibitem [{\citenamefont {Venkatesan}\ \emph {et~al.}(2008)\citenamefont
  {Venkatesan}, \citenamefont {Vlooswijk}, \citenamefont {Kooi}, \citenamefont
  {Morelli}, \citenamefont {Palasantzas}, \citenamefont {De~Hosson},\ and\
  \citenamefont {Noheda}}]{venkatesan2008monodomain}%
  \BibitemOpen
  \bibfield  {author} {\bibinfo {author} {\bibfnamefont {S.}~\bibnamefont
  {Venkatesan}}, \bibinfo {author} {\bibfnamefont {A.}~\bibnamefont
  {Vlooswijk}}, \bibinfo {author} {\bibfnamefont {B.~J.}\ \bibnamefont {Kooi}},
  \bibinfo {author} {\bibfnamefont {A.}~\bibnamefont {Morelli}}, \bibinfo
  {author} {\bibfnamefont {G.}~\bibnamefont {Palasantzas}}, \bibinfo {author}
  {\bibfnamefont {J.~T.}\ \bibnamefont {De~Hosson}}, \ and\ \bibinfo {author}
  {\bibfnamefont {B.}~\bibnamefont {Noheda}},\ }\href@noop {} {\bibfield
  {journal} {\bibinfo  {journal} {Phys. Rev. B}\ }\textbf {\bibinfo {volume}
  {78}},\ \bibinfo {pages} {104112} (\bibinfo {year} {2008})}\BibitemShut
  {NoStop}%
\bibitem [{\citenamefont {Haun}\ \emph {et~al.}(1987)\citenamefont {Haun},
  \citenamefont {Furman}, \citenamefont {Jang}, \citenamefont {McKinstry},\
  and\ \citenamefont {Cross}}]{haun1987thermodynamic}%
  \BibitemOpen
  \bibfield  {author} {\bibinfo {author} {\bibfnamefont {M.~J.}\ \bibnamefont
  {Haun}}, \bibinfo {author} {\bibfnamefont {E.}~\bibnamefont {Furman}},
  \bibinfo {author} {\bibfnamefont {S.}~\bibnamefont {Jang}}, \bibinfo {author}
  {\bibfnamefont {H.}~\bibnamefont {McKinstry}}, \ and\ \bibinfo {author}
  {\bibfnamefont {L.}~\bibnamefont {Cross}},\ }\href@noop {} {\bibfield
  {journal} {\bibinfo  {journal} {J. Appl. Phys.}\ }\textbf {\bibinfo {volume}
  {62}},\ \bibinfo {pages} {3331} (\bibinfo {year} {1987})}\BibitemShut
  {NoStop}%
\end{thebibliography}
\end{document}